\documentclass[aps,prb,twocolumn,groupedaddress]{revtex4-2}

\usepackage{amsmath}
\usepackage{amssymb}
\usepackage{amsthm}
\usepackage{mathtools}
\usepackage{hyperref}

\usepackage{color}
\usepackage{import}
\usepackage{braket}

\begin{document}


\title{Photoinduced prethermal order parameter dynamics in the two-dimensional large-$N$ Hubbard-Heisenberg model}


\author{Alexander Osterkorn}
\email[]{osterkorn@theorie.physik.uni-goettingen.de}

\author{Stefan Kehrein}

\affiliation{Institute for Theoretical Physics,  Georg-August-Universit\"at G\"ottingen, Friedrich-Hund-Platz 1 - 37077  G\"ottingen, Germany}


\date{\today}

\begin{abstract}
We study the microscopic dynamics of competing ordered phases in a two-dimensional correlated electron model, which is driven with a pulsed electric field of finite duration.
In order to go beyond a mean-field treatment of the electronic interactions we adopt a large-$N$ generalization of the Hubbard model and combine it with the semiclassical fermionic truncated Wigner approximation as a time evolution method.
This allows us to calculate dephasing corrections to the mean-field dynamics and to obtain stationary states, which we interpret as prethermal order.
We use this framework to simulate the light-induced transition between two competing phases (bond density wave and staggered flux) and find that the post-pulse stationary state order parameter values are not determined alone by the amount of absorbed energy but depend explicitly on the driving frequency and field direction.
While the transition between the two prethermal phases takes place at similar total energies in the low- and high-frequency regimes, we identify an intermediate frequency regime for which it occurs with minimal heating.
\end{abstract}


\maketitle

\section{Introduction\label{sec:introduction}}

The interaction of electromagnetic fields with matter is at the heart of many measurement techniques, like angle-resolved photoemission~\cite{Damascelli2003,Reinert2005}, that have shaped our today's understanding of strongly correlated electron materials in equilibrium.
Ultrashort laser pulses~\cite{Diels2006} allow to transfer the setups to non-equilibrium situations where, for instance, the dynamics of electrons under strong laser driving can be recorded in real time.
The list of reported genuine out-of-equilibrium phenomena in driven solids includes the transient manipulation of band structures (``Floquet engineering''~\cite{Bukov2015,Oka2019}), the switching to hidden states in materials~\cite{Stojchevska2014} and light-induced superconductivity \cite{Fausti2011,Mitrano2016,Paeckel2020}.
A central research question is to identify ``non-thermal pathways''~\cite{delaTorre2021} to control ordered phases in materials.
Currently, there is a particular interest in (quasi-) two-dimensional materials like transition metal dichalcogenides~\cite{Manzeli2017} that may allow for important applications.
Correlated electron systems typically host a number of competing ordering tendencies, which can often be captured already at mean-field level~\cite{Sau2014,Laughlin2014}.
Selecting and enhancing particular types of order, e.g. superconductivity~\cite{Fu2014,Raines2015,Patel2016,Sentef2016,Sentef2017}, with electromagnetic fields presents the prospect of (transiently) engineering desired physical properties in materials.

On the theory side, such scenarios can often be described within phenomenological time-dependent Ginzburg-Landau theories~\cite{Dolgirev2020,Sun2020,Sun2020a,Grandi2021}.
Still, microscopic ``bottom-up'' modeling allows for a more systematic inclusion of electronic correlation effects and can be made more material-specific.
Despite a lot of ongoing method development in one~\cite{Schollwock2011,Paeckel2019} and high spatial dimensions~\cite{Aoki2014},
it is challenging to implement such simulations due to the lack of a general purpose numerical time evolution method, especially in two spatial dimensions.
Non-equilibrium Green's functions~\cite{Schlunzen2019} are the most versatile scheme, but they are, despite recent progress~\cite{Schlunzen2020,Joost2020,Kaye2021,Stahl2022}, computationally demanding and often rely on additional approximations like the generalized Kadanoff-Baym ansatz~\cite{Lipavsky1986,Karlsson2018}.
Infinite periodic driving can be described by effective Floquet Hamiltonians~\cite{Bukov2015,Eckardt2017,Vogl2019,Vogl2020b} but the effect of heating~\cite{Mallayya2019,Ikeda2021} and possibly finite pulse durations~\cite{Novicenko2017,Novicenko2022} restrict their applicability.


In this work we consider a two-dimensional lattice system driven with a pulsed electric field of finite duration as it is used in time-resolved ARPES experiments \cite{Freericks2009,Freericks2015}.
In order to work within a well-controlled theoretical framework,
we make use of the limit of large fermion degeneracy.
In such a so-called large-$N$ model, the two electronic spin states are generalized to $N$ internal ``flavor'' degrees of freedom.
$N \rightarrow \infty$ is a natural classical limit~\cite{Yaffe1982} and one can derive a systematic $1/N$ expansion around it~\cite{Marston1989,Sachdev1991,Read1991}.
We choose the two-dimensional large-$N$ Hubbard-Heisenberg model~\cite{Affleck1988,Marston1989} as a paradigmatic SU($N$)-generalization of the Hubbard model with two competing phases, i.e. bond density wave and staggered flux order.
SU($N$)-symmetric models have been and continue to be a popular topic of theoretical research \cite{Yanatori2016,Lee2018,Ibarra-Garcia-Padilla2021,Yoshida2022}, in particular in one spatial dimension~\cite{Zhao2007,Buchta2007,Assaraf1999,Szirmai2008}.
An additional motivation stems from experiments with ultracold atoms \cite{Honerkamp2004,Manmana2011,Hofrichter2016}.

The main focus of our study is the formation of prethermal order subsequent to the pulse.
This refers to a regime before thermalization, when quantum systems can dephase into quasi-stationary states with long lifetimes \cite{Moeckel2008,Eckstein2009,Herrmann2017,Alexander2022},
which are characterized by the existence of additional conserved quantities.
This prethermalization dynamics will depend on the specifics of the drive \cite{Alexander2022} and offers -- if sufficiently understood -- a way to transiently manipulate ordered phases in a material or even to induce order which is inaccessible in equilibrium.
In this sense prethermal order is particularly interesting because -- unlike thermal states -- it is not only determined by the total amount of absorbed energy.
Methodwise, we combine the fermionic truncated Wigner approximation (fTWA) \cite{Lacroix2014,Davidson2017} with the large-$N$ setup allowing us to include dephasing corrections at order $1/N$ to the classical $N \rightarrow \infty$ dynamics, which are responsible for the formation of prethermal order~\cite{Osterkorn2020}.

The text is structured as follows:
We start by introducing the model with its mean-field phases and the time evolution scheme in Sec.~\ref{sec:methods}.
In particular, we describe how we obtain stationary initial states at finite $N$.
In Sec.~\ref{sec:photo} we discuss the driven system, the effect of dephasing and how drive parameters influence the transition between the competing phases.
Sec.~\ref{sec:discussion} contains a summary as well as a discussion of the results and possible next steps.

\section{Model and methods\label{sec:methods}}

\subsection{The large-$N$ Hubbard-Heisenberg model and its classical limit}

\begin{figure}
 \resizebox{0.45\textwidth}{!}{\Large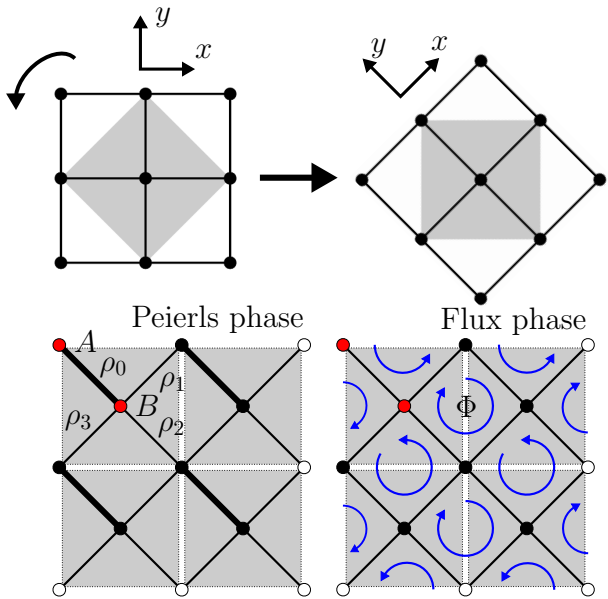}
 \caption{Upper row: Unit cell used in this paper with two atoms per cell. Lower row: Sketch of the spatial structure of the two competing equilibrium phases Peierls (one strong bond $\rho_0$ and three identical weak bonds $\rho_{1,2,3}$ per unit cell) and Flux (four complex bonds with equal magnitude and non-vanishing plaquette flux).
 \label{fig:tilted_unit_cell}}
\end{figure}

We consider fermionic annihilation and creation operators $c_{i\alpha}^{(\dagger)}$ with lattice site $i$ and $\alpha = 1, \dots, N$ internal flavor states.
$N = 2$ corresponds to the standard spin-full fermions.
The Hamiltonian of the SU($N$)-symmetric Hubbard-Heisenberg model reads as follows:
\begin{align} \begin{split}
 \hat H &= t_\text{h} \sum_{\langle i, j \rangle} \left( \sum_{\alpha=1}^N c_{i\alpha}^\dagger c_{j\alpha} + H. c. \right) \\
 &- \frac{J}{N} \sum_{\langle i, j \rangle} \bigg| \sum_{\alpha = 1}^N c_{i\alpha}^\dagger c_{j\alpha} \bigg|^2 \\
 &+ \frac{U}{N} \sum_i \left( \sum_{\alpha=1}^N c_{i\alpha}^\dagger c_{i\alpha} - \frac{N}{2} \right)^2
 \label{eq:hamiltonian_cs}
\end{split} \end{align}
$t_\text{h(op)}$, $J$ and $U$ are free parameters.
Here we have kept Marston's and Affleck's original sign convention without a minus sign in front of the hopping term.
The $U$-term describes a Hubbard-type interaction and the $J$-term derives from a Heisenberg interaction.
Both interaction types are included explicitly in the model to allow for a smooth interpolation between the limits $N \rightarrow \infty$ and $U \rightarrow \infty$ \cite{Marston1989}.

One can straightforwardly reformulate the model in terms of the operators
\begin{equation}
 \hat \rho_{i\alpha,j\beta} =  c_{i\alpha}^\dagger c_{j\beta} - \frac{1}{2} \delta_{ij} \delta_{\alpha\beta} ,
\end{equation}
similar to the one-particle reduced density matrix.
Going one step further we introduce flavor-averaged operators
\begin{equation}
 \hat \rho_{ij} = \frac{1}{N} \sum_\alpha \hat \rho_{i\alpha, j\alpha},
\end{equation}
which obey the commutation relations
\begin{equation}
 [ \hat \rho_{ij}, \hat \rho_{mn} ] = \frac{1}{N} ( \delta_{jm} \hat \rho_{in} - \delta_{in} \hat \rho_{mj} ) .
\end{equation}
We can readily see that $\frac{1}{N}$ plays the role of an effective $\hbar$ and thus of the semiclassical expansion parameter.
Hence, the classical limit corresponds to $N \rightarrow \infty$ and the operators $\hat \rho_{ij}$ become commuting classical variables $\rho_{ij}$.

At finite $N$, the ordering of products of the operators $\hat \rho_{ij}$ matters in principle and can lead to different classical limits.
The most common ordering convention for two quantum operators $\hat A$, $\hat B$ is symmetrization $\frac{1}{2}\big( \hat A \hat B + \hat B \hat A \big)$ leading to the Wigner-Weyl framework.
In fact, the Hamiltonian \eqref{eq:hamiltonian_cs} is not modified by the symmetrization since  $\frac{1}{2} \big( \hat \rho_{i j} \hat \rho_{j i} +  \hat \rho_{j i} \hat \rho_{i j} \big) = \hat \rho_{i j} \hat \rho_{j i} + \frac{1}{2 N} \big( \hat \rho_{ii} - \hat \rho_{jj} \big)$.
Here, the sum over nearest neighbor pairs yields a cancellation of the order $1/N$ terms and so the classical limit of our model reads
\begin{align} \begin{split}
 H = N \bigg\{ &\sum_{\langle ij \rangle} \left[ t_\text{h} (\rho_{ij} + \rho_{ji}) - J \left| \rho_{ij} \right|^2 \right] \\
 + U &\sum_i \rho_{ii}^2 \bigg\} .
 \label{eq:ftwa_hamiltonian}
\end{split} \end{align}
One can also view this classical Hamiltonian as obtained from a mean-field decoupling of the original interaction terms.
The global prefactor $N$ in front gives rise to a large deviation form $\text{e}^{-N \beta \hat h}$ of the density operator that leads to a suppression of fluctuations as $N \rightarrow \infty$.

\subsection{Equations of motion and fTWA}

For the time evolution we adopt a variant of the semiclassical truncated Wigner approximation (TWA) scheme \cite{Polkovnikov2010}, the fermionic TWA (fTWA \cite{Davidson2017,Schmitt2020,Sajna2020,Osterkorn2020}).
The method was independently developed earlier under the name ``stochastic mean-field approach'' \cite{Lacroix2014,Ulgen2019,Czuba2020,Schroedter2022}.
It is based on the mean-field/classical equations of motion for the operators $\hat\rho_{ij}$.
These can be obtained, for instance, by calculating the Heisenberg equations of motion of the quantum Hamiltonian and then sending $N \rightarrow \infty$.
Equivalently, one can start from the classical Hamiltonian  and derive its Hamiltonian equations of motion which, dressed by an additional factor of $i$, coincide with the Heisenberg equations of motion.
They read:
\begin{align} \begin{split}
 i \partial_t \rho_{ij} = &\sum_{a(j)} (t_\text{h} - J \rho_{a(j), j}) \rho_{i, a(j)} \\
 - &\sum_{a(i)} (t_\text{h} - J \rho_{i, a(i)}) \rho_{a(i), j} \\
 + &2 U (\rho_{jj} - \rho_{ii}) \rho_{ij}
 \label{eq:hubhei_eom}
\end{split} \end{align}
$a(i)$ denotes the set of all nearest neighbor sites of site $i$.
In all TWA methods quantum mechanical expectation values of operators are calculated from averages over classical Hamiltonian trajectories
whose initial values are sampled from a (quasi-)propability distribution, the Wigner function $W$.
Within fTWA the operators $\hat\rho_{i\alpha,j\beta}$ act as the dynamical variables.
Here, we apply the scheme directly to the SU($N$) variables $\rho_{ij}$ \cite{Osterkorn2020}.
The expectation value of a quantum operator $\hat O[\hat \rho_{ij}]$ that can be expressed in terms of some $\rho_{ij}$ reads
\begin{equation}
 \big\langle \hat O(t) \big\rangle = \int \text{d} \rho_{ij}(0) W \big( \rho_{ij}(0) \big) O_\text{W} \big( \rho_{ij}(t) \big) ,
\end{equation}
where $O_\text{W}(\rho_{ij})$ is the so-called Weyl symbol of $\hat O$.
It can be different for different ways of writing $\hat O$ in terms of the $\hat\rho_{ij}$ \cite{Sajna2020}.
In the following we would like to calculate phase angles and absolute values of $\hat\rho_{ij}$, i.e.
operations whose Weyl symbols are not obvious.
However, in the classical theory $N \rightarrow \infty$ it is clear that one can calculate these observables straightforwardly from the values of the $\rho_{ij}$.
Analogous to $(\hat x \hat p)_\text{W} = x p + \mathcal{O}(\hbar)$ in single-particle TWA,
corrections to Weyl symbols will be of order $\frac{1}{N}$,
e.g. $| \bullet |_W = | \bullet | + \mathcal{O}(\frac{1}{N})$.
Since we will work at large $N$,
we are justified to neglect these corrections.
The Wigner function is constructed as a multivariate Gaussian reproducing the quantum mechanical means and (connected) covariances of the $\hat \rho_{ij}$ in a quantum state $\ket{\psi}$.
\begin{align} \begin{split}
 \big\langle \rho_{ij} \big\rangle_W &\overset{!}{=} \big\langle \hat\rho_{ij} \big\rangle_\psi, \\
 \big\langle \rho_{ij} \rho_{mn} \big\rangle_W^\text{cc} &\overset{!}{=} \frac{1}{2} \big\langle \hat \rho_{ij} \hat\rho_{mn} + \hat\rho_{mn} \hat\rho_{ij} \big\rangle_\psi^\text{cc} \sim \frac{1}{N}
 \label{eq:wignerfunc_relations}
\end{split} \end{align}
The fTWA method generates a hierarchy of correlations akin to the BBGKY hierarchy~\cite{Bonitz2016}.
Despite some recent discussion~\cite{Czuba2020} about the connection of the two, a systematic picture how to formulate fTWA in terms of more conventional hierarchy approximations is still missing.

\subsection{Equilibrium states of the model}

\begin{figure}
	\includegraphics[width=0.5\textwidth]{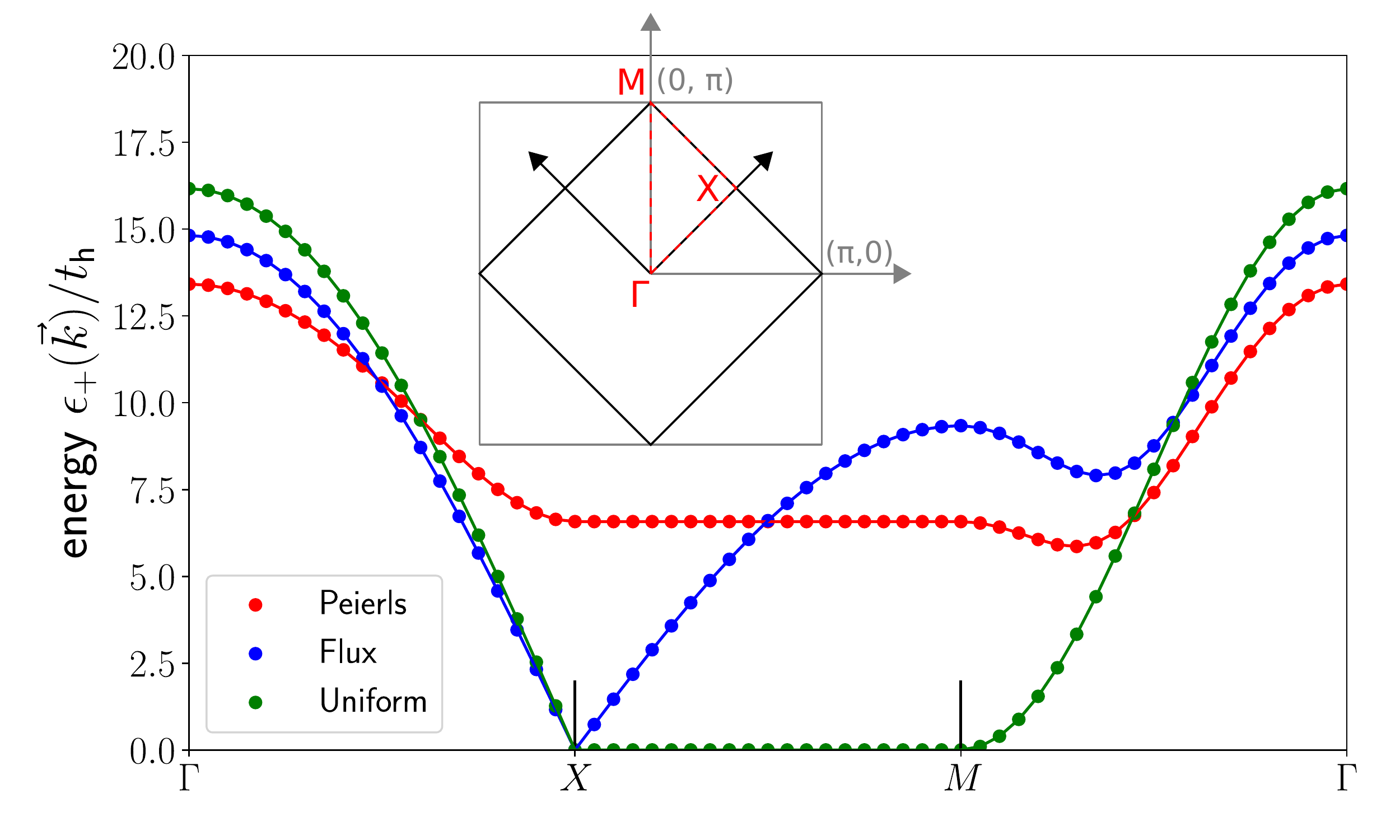}
	\caption{Single-particle mean-field band structure for
		$J/t_\text{h} = 15$ and a system size of $V_u = 41 \times 41$ unit cells.
		The spectrum is symmetric around energy zero; here we only plot the upper half with positive energy.
		The Peierls phase is gapped, while the Flux phase is gapless.
		Inset: Sketch of the reduced Brillouin zone derived from the unit cell in Fig.~\ref{fig:tilted_unit_cell}.\label{fig:brillouin_disp}}
\end{figure}

\begin{figure}
	\includegraphics[width=0.5\textwidth]{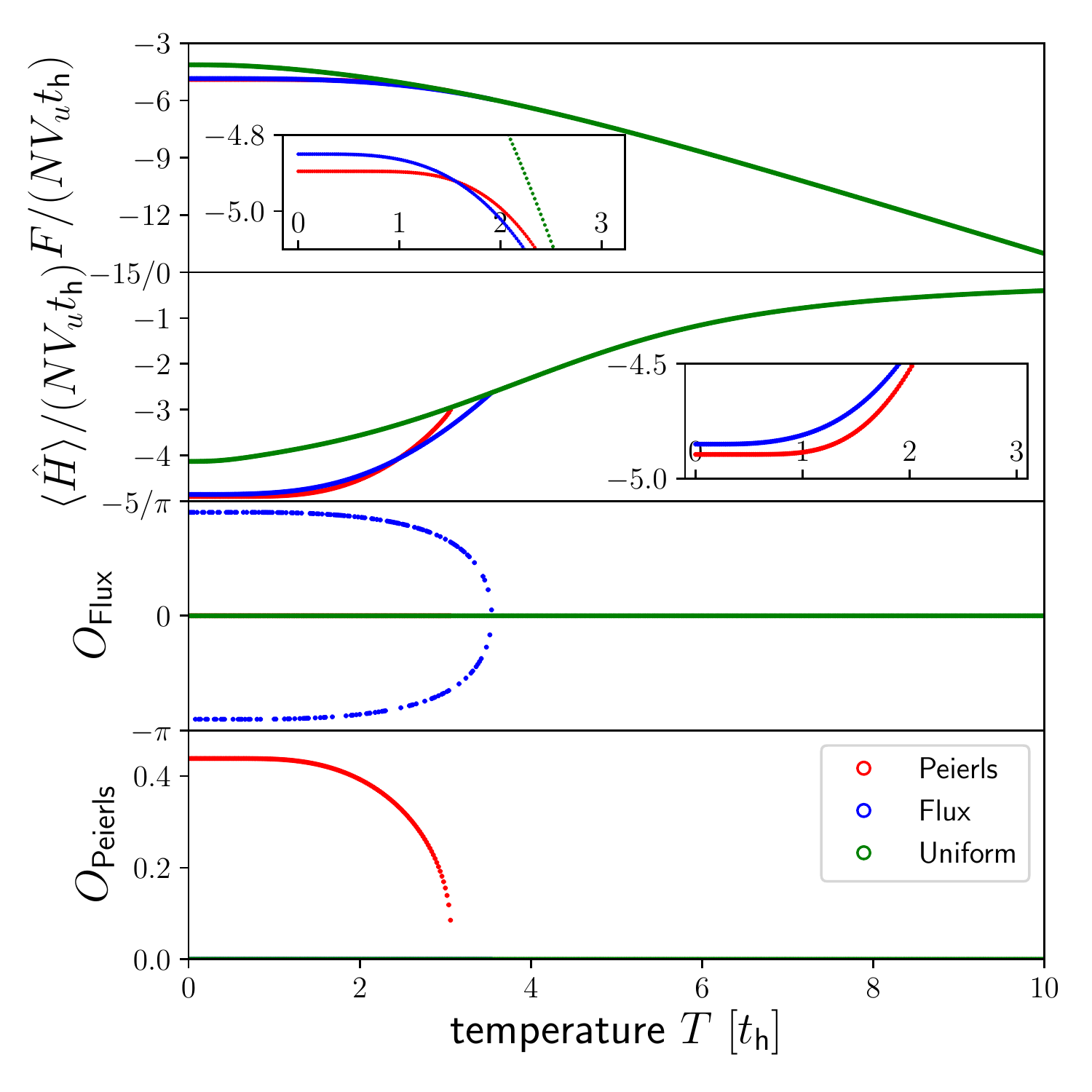}
	\caption{Finite-temperature equilibrium states for the $N \rightarrow \infty$ Hubbard-Heisenberg model with $V_u = 11 \times 11$ unit cells, $J/t_\text{h} = 15$ and $U = 0$.
		Upper two rows: Helmholtz free energy $F$ and total (internal) energy $\langle \hat H \rangle$ per unit cell.
		Lower two rows: Order parameters for the Flux and Peierls phases displaying the typical mean-field scaling.
		At temperature $T \approx 1.5 t_\text{h}$ there is a first order transition from Peierls to Flux order.
		\label{fig:eq_phases_temp_J15}}
\end{figure}

The $N \rightarrow \infty$ equilibrium zero temperature phase diagram was first obtained using field theoretical methods \cite{Affleck1988,Marston1989}.
Later it was re-examined \cite{Assaad2005} with quantum Monte-Carlo also for finite $N < \infty$.
For a self-consistent mean-field solution,
one needs to define a unit cell.
We choose Marston's and Affleck's original tilted two-site unit cell depicted in the top row of Fig.~\ref{fig:tilted_unit_cell}.
We will always use periodic boundary conditions (with respect to the lattice of unit cells) for quadratic systems with at least $11 \times 11$ cells, i.e. $V = 2 \cdot 11^2 = 242$ lattice sites.
The choice of the unit cell corresponds to a reduced Brillouin zone, which is sketched with black color in Fig.~\ref{fig:brillouin_disp}.

\subsubsection{Mean-field solution at $N \rightarrow \infty$}


The limit $N \rightarrow \infty$ is a mean-field limit that corresponds to a Hartree decoupling of the interaction:
The way we write the interaction terms with $\hat\rho_{ij}$ operators determines their decoupling.
Leaving the details for Appendix~\ref{app:details_mf},
we obtain a set of single-particle eigenenergies $E_{k\pm} = \pm \big( \big| \epsilon_k - \chi_k \big|^2 + U^2 (\rho_B - \rho_A)^2 \big)^{1/2}$.
Charge-density wave (CDW) states $\rho_A \neq \rho_B$ would lead to a gap in the single-particle spectrum.
Some caution needs to be taken when looking for a saddle point in the mean-field free energy.
The saddle point curvature in the $U$-direction is inverse to one of the $J$-term (some more details are given in the original publication \cite{Marston1989}).
The internal energy (expectation value of the Hamiltonian) per flavor is given by
\begin{align} \begin{split}
 H/N &= \frac{V}{2} \Big( - U (\rho_A^2 + \rho_B^2) + J \sum_{i = 0}^3 | \rho_0 |^2 \Big) \\
 &\quad + \sum_k E_k \left( \rho_{k+,k+} - \rho_{k-,k-} \right) ,
 \label{eq:mf_ham_with_U}
\end{split} \end{align}
where $V$ is the total number of lattice sites.
Using
\begin{equation}
 U \big( \rho_A^2 + \rho_B^2 \big) = \frac{U}{2} \big( \rho_A + \rho_B \big)^2 + \frac{U}{2} \big( \rho_A - \rho_B \big)^2 ,
\end{equation}
where $\rho_A + \rho_B$ is fixed by the filling,
we see that $\rho_A = \rho_B$ is required for a stable saddle point.
It implies that in this model CDWs are thermodynamically unfavorable due to the $U$-term in the Hamiltonian.
In particular, the value of $U$ is irrelevant for the mean-field phase diagram.
It is worth noting that this is different for nearest-neighbor density-density interactions instead of the Hubbard interaction.
Setting $\rho_A = \rho_B$ we obtain
\begin{align} \begin{split}
 H/N &= \frac{V}{2} J \big( | \rho_0 |^2 + | \rho_1 |^2 + | \rho_2 |^2 + | \rho_3 |^2 \big) \\
 &\qquad + \sum_k E_k \left( \rho_{k+,k+} - \rho_{k-,k-} \right), \\
 E_k &= 2t_\text{h} \big( \cos(k_x) + \cos(k_y) \big) \\
 &\quad - J \big( \rho_0^\ast \text{e}^{-i k_y} + \rho_1^\ast \text{e}^{i k_x} \\
 &\qquad\quad + \rho_2 \text{e}^{i k_y} + \rho_3 \text{e}^{-i k_x} \big) .
 \label{eq:mf_ham}
\end{split} \end{align}

Due to the symmetry of $E_{k\pm}$ around $E = 0$, the ground state at half filling is always obtained by setting $\rho_{k-,k-} = \frac{1}{2}$ and $\rho_{k+,k+} = -\frac{1}{2}$ for all momenta $k$.
We find self-consistent values for $\rho_0$, $\rho_1$ at  zero temperature from the numerical minimization of \eqref{eq:mf_ham} using a simulated annealing algorithm.

At half filling three phases are realized.
Firstly, the dimerized bond density wave \emph{Peierls phase} at large $J/t_\text{h}$ has got real $\rho_i$ with one strong bond $\rho_0 \gg \rho_1 = \rho_2 = \rho_3$.
We introduce the following order parameter for the Peierls phase:
\begin{equation}
 O_\text{Peierls} = | \rho_0 | - \max_{i = 1,2,3} | \rho_i | .
\end{equation}
Secondly, the \emph{staggered Flux phase} (also known as DDW phase) at intermediate $J/t_\text{h}$ comes with complex bonds $\rho_i$ that are equal in magnitude and multiply to a flux operator $\Pi = \Pi_0 \text{e}^{i \Phi}$ around a plaquette with non-vanishing phase $\Phi \neq 0$.
Hence, we will consider
\begin{equation}
 O_\text{Flux} = \Phi .
\end{equation}
Lastly, there is the \emph{uniform phase} which is a stable saddle point at all values of $J$ but has the lowest free energy only at $J = 0$.
In the uniform phase the unit cell bonds are real and $O_\text{Peierls} = O_\text{Flux} = 0$.
The single-particle band structures obtained from the mean-field calculation are plotted in Fig.~\ref{fig:brillouin_disp} for the reduced Brillouin zone.
The bands in the Figure correspond to the electron removal/addition energies and do not directly contain information about excited states of the system.

\subsubsection{Non-zero temperatures}

Since we are interested in the values of the order parameters after energy absorption due to the driving,
we would like to compare our results with equilibrium at temperatures $T > 0$.
In order to determine the finite-temperature states, we need to include the entropy term in the free energy.
Since on the mean-field level the ground state may be understood as a simple product state of a single fermion species
\begin{equation}
 \ket{\tilde{\Psi}_0^\text{MF}} = \prod_{\epsilon_{k\pm} < \epsilon_F} c_{k\pm}^\dagger \ket{0}
\end{equation}
with respect to the mean-field basis,
we can use the formula for the entropy of free fermions
\begin{align} \begin{split}
S/N = \frac{1}{T} \sum_{\epsilon_k} &\bigg[ n_\text{FD}(\epsilon_k) \big( \epsilon_k - \mu \big) \\
&\quad + T \ln \Big( 1 + \text{e}^{-(\epsilon_k - \mu)/T} \Big) \bigg]
\label{eq:entropy_free_fermions}
\end{split} \end{align}
with the Fermi-Dirac distribution $n_\text{FD}(\epsilon)$.
This yields the following (Helmholtz) free energy $F = \langle H \rangle - T S$,
\begin{align} \begin{split}
 F/N &= J \sum_{\langle ij \rangle} \left| \rho_{ij} \right|^2 - \sum_{\epsilon_k} \beta^{-1} \ln \Big( 1 + \text{e}^{-\beta(\epsilon_k - \mu)} \Big) .
 \label{eq:free_en_temp}
\end{split} \end{align}

We determined the minima of the free energy as well using simulated annealing.
In this paper we concentrate on half filling and on a value of $J = 15 t_\text{h}$, which lies in a region of the phase diagram where the Peierls phase has the lowest free energy.
The results are shown in Fig.~\ref{fig:eq_phases_temp_J15}.
From temperature zero up to $T \approx 3.5 t_\text{h}$ there is a regime in which all three phases are stable.
Up to $T \approx 1.6 t_\text{h}$ the Peierls phase has the lowest energy, at more elevated temperatures the flux phase is preferred.
In the high-temperature phase all order is destroyed and only the uniform phase is left.
The order parameters display the typical mean-field scaling with exponent $\beta \sim \frac{1}{2}$.
For larger values of $J$ the range of temperatures with stable order beyond the uniform phase will be broader.

\subsubsection{Order parameters in the finite-$N$ model}

It was demonstrated using quantum Monte Carlo~\cite{Assaad2005} that the mean-field saddle points yield a qualitatively correct picture of the finite-$N$ Hubbard-Heisenberg model down to $N = 6$, albeit with renormalized order parameters and phase boundaries.
We can therefore work with the same order parameter definitions as for the infinite-$N$ model but we need to take care how to consistently calculate them within the TWA scheme.
In the numerical determination of the Flux order parameter we average $e^{i \Phi}$ over all trajectories, i.e. we work with numbers on the unit circle.
However, one needs to take care of the spontaneous breaking of the orientation symmetry of the plaquette flux.
There are two ergodic components, for which $\Phi$ lies in the intervals $[0, \pi]$ and $[-\pi,0]$, respectively.
Different fTWA trajectories can select different ergodic components such that direct averaging of the order parameter may become unphysical (analogous to averaging the magnetization of a ferromagnet without explicit symmetry breaking).
Hence, in order to calculate the Flux order parameter directly, we introduce a weak symmetry breaking $i \epsilon_{m n} \big( \hat j_{mn} - \hat j_{nm} \big)$ that selects the $\Phi \in [0,\pi]$ component.
The current operator (for lattice sites $m,n$) is given by
\begin{equation}
\hat j_{mn} = i t_\text{h} \big( c_m^\dagger c_n - c_n^\dagger c_m \big) = -2 t_\text{h} \operatorname{Im}( \hat\rho_{m n} ) .
\end{equation}
We tried out a few symmetry breaking strengths and concluded that $\epsilon \sim 10^{-3}$ gives the best balance of breaking the symmetry but not influencing the dynamics too much.
Another probe for long-range Flux order which is well defined without symmetry breaking field~\cite{Schollwock2003} are current-current correlators of the form $\big\langle \hat j_{mn} \hat j_{rs} \big\rangle$.
We calculate these for two bonds at the largest spatial separation in the system to detect long-range order.
Making use of the translational symmetry of the system we determine the correlators for all translationally equivalent pairs of unit cells and average over them.

\subsubsection{Initial state for finite $N$}

\begin{figure}
    \centering
    \includegraphics[width=0.5\textwidth]{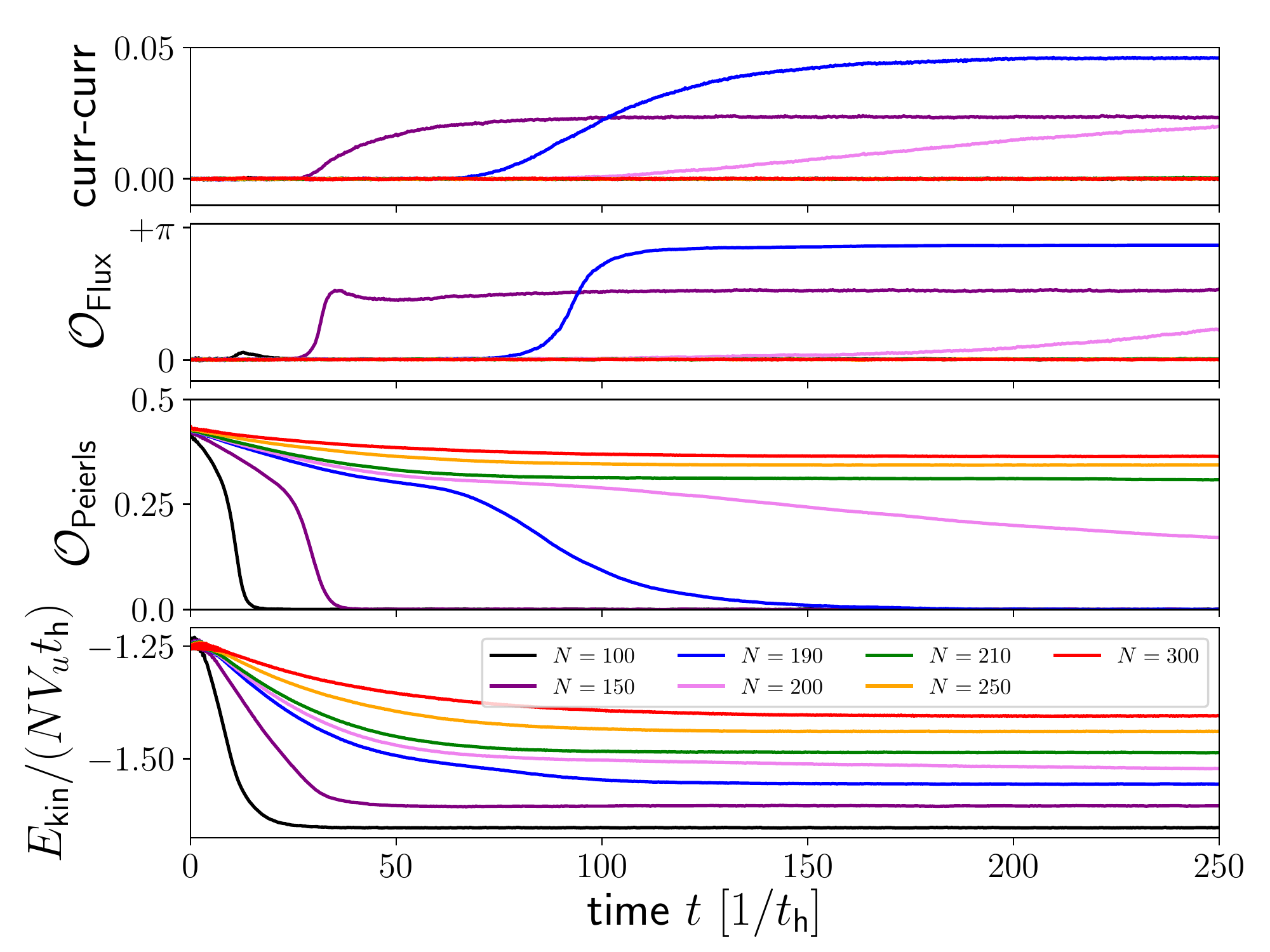}
    \caption{Initial relaxation dynamics generated by a Gaussian Wigner function corresponding to the state \eqref{eq:ini_state_fin_N} in the finite-$N$ Hubbard-Heisenberg model with $J = 15 t_\text{h}$, $U = 0$ and $V_u = 11 \times 11$ unit cells. The order parameters and the kinetic energy reach a stationary state before time $t = 250 t_\text{h}^{-1}$ for all shown values of $N$ except for $N = 200$ for which the relaxation takes longer than $1000 t_\text{h}^{-1}$.
    \label{fig:stat_order_params_vs_time_mult_N}}
\end{figure}

\begin{figure}
 \includegraphics[width=0.5\textwidth]{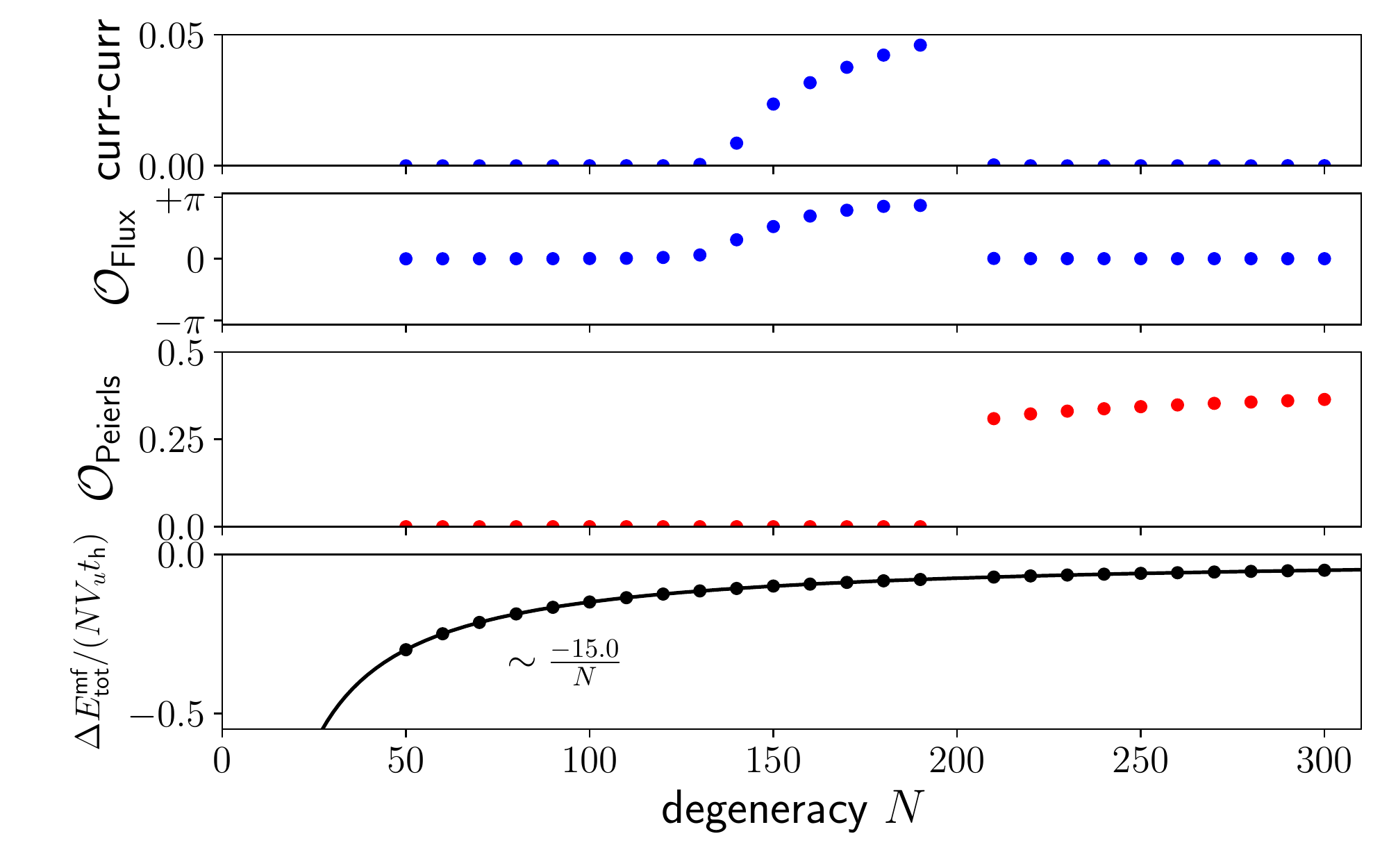}
 \caption{Stationary values of the order parameters after the warm-up shown in Fig.~\ref{fig:stat_order_params_vs_time_mult_N}.
 $\Delta E_\text{tot}^\text{mf} = E_\text{tot} - E_\text{tot}^\text{mf}$ measures the amount of correlation energy that is added to the system at time $t = 0$.
 We omit the data at $N = 200$ because they were not relaxed after time $1000 t_\text{h}^{-1}$.
 \label{fig:stat_order_params_vs_N}}
\end{figure}

For $N < \infty$, it would be desirable to explicitly construct the equilibrium states of the model as controlled starting points for the light-driven dynamics.
In complete generality, this requires the calculation of $1/N$ corrections to the saddle points and is a non-trivial task.
The equilibrium states so obtained would be stationary under the semiclassical time-evolution.
To proceed without a full calculation, we resort to a simpler strategy and make direct use of the fTWA dynamics.
This is possible in at least to ways: We can adiabatically switch on the interactions or we can construct an approximate Wigner function which relaxes to a stationary state in the phase of interest.
The first strategy has the advantage that for sufficiently slow switching the system can be prepared in an equilibrium ground state, i.e. without additional heating.
However, it turns out to be challenging to implement this in practice.
For switching times of $1000 t_\text{h}^{-1}$ the final state energies we obtained were significantly higher than those from the strategy discussed below.
Another difficulty is posed by the fact that all three phases are stable and a suitable transient symmetry breaking might be required to reach the desired phase.
We leave for future work a more detailed discussion of adiabatic switching with fTWA.

In the following, we employ the conceptually simpler second strategy and prepare a non-stationary Gaussian Wigner function \eqref{eq:wignerfunc_relations}, which derives from an initial product state at half filling of the form
\begin{equation}
 \ket{\Psi_0^N} = \prod_k \prod_{\alpha = 1}^N c_{k-,\alpha}^\dagger \ket{0} ,
 \label{eq:ini_state_fin_N}
\end{equation}
where $k\pm$ label the mean-field modes.
The idea behind this is that we take the self-consistent mean-field data for the one-particle density matrix and assume it will be close to the equilibrium at large but finite $N$.
The classical phase space includes all variables $\rho_{k\pm, l\pm}$ for momenta $k, l$.
The non-vanishing covariances of the state \eqref{eq:ini_state_fin_N} are of the form $\langle \rho_{\alpha\beta} \rho_{\alpha \beta} \rangle$ and $\langle \rho_{\alpha\beta} \rho_{\beta \alpha} \rangle$.
The complete initial data for a Gaussian Wigner function of this type is listed in Appendix~\ref{app:wigner}.
The undriven system will relax to a stationary state on a characteristic warm-up time scale.
Later, when we investigate the dynamics of the driven system we switch the drive on only after this warm-up.
A disadvantage of this procedure is that the resulting stationary state is not a ground state.
This, however, is not too problematic for this paper since we are only interested in a stationary reference state in the Peierls phase.
During the warm-up the system may get pushed out of the Peierls phase thereby restricting the range of $N$ to which this strategy can be applied.
Fig.~\ref{fig:stat_order_params_vs_time_mult_N} shows the time evolution of the order parameters starting from a Wigner function derived from \eqref{eq:ini_state_fin_N}.
For $N \gtrapprox 200$ the dynamics leads to a decrease of the Peierls order parameter, while
$\mathcal{O}_\text{Flux}$ remains zero at all times.
Around $N \approx 200$ the system transitions to the Flux phase and the dynamics is very slow.
For smaller values $N \lessapprox 150$,
the system ends up in the Uniform phase.
In Fig.~\ref{fig:stat_order_params_vs_N}
we collect the stationary values of the order parameters in Fig.~\ref{fig:stat_order_params_vs_time_mult_N}.
We conclude that for $J = 15 t_\text{h}$ and a lattice with $11 \times 11$ unit cells we can use this approximative approach for the initial Wigner function down to values of about $N = 200$.

\section{Photoexcitations\label{sec:photo}}

\begin{figure}
 \includegraphics[width=0.48\textwidth]{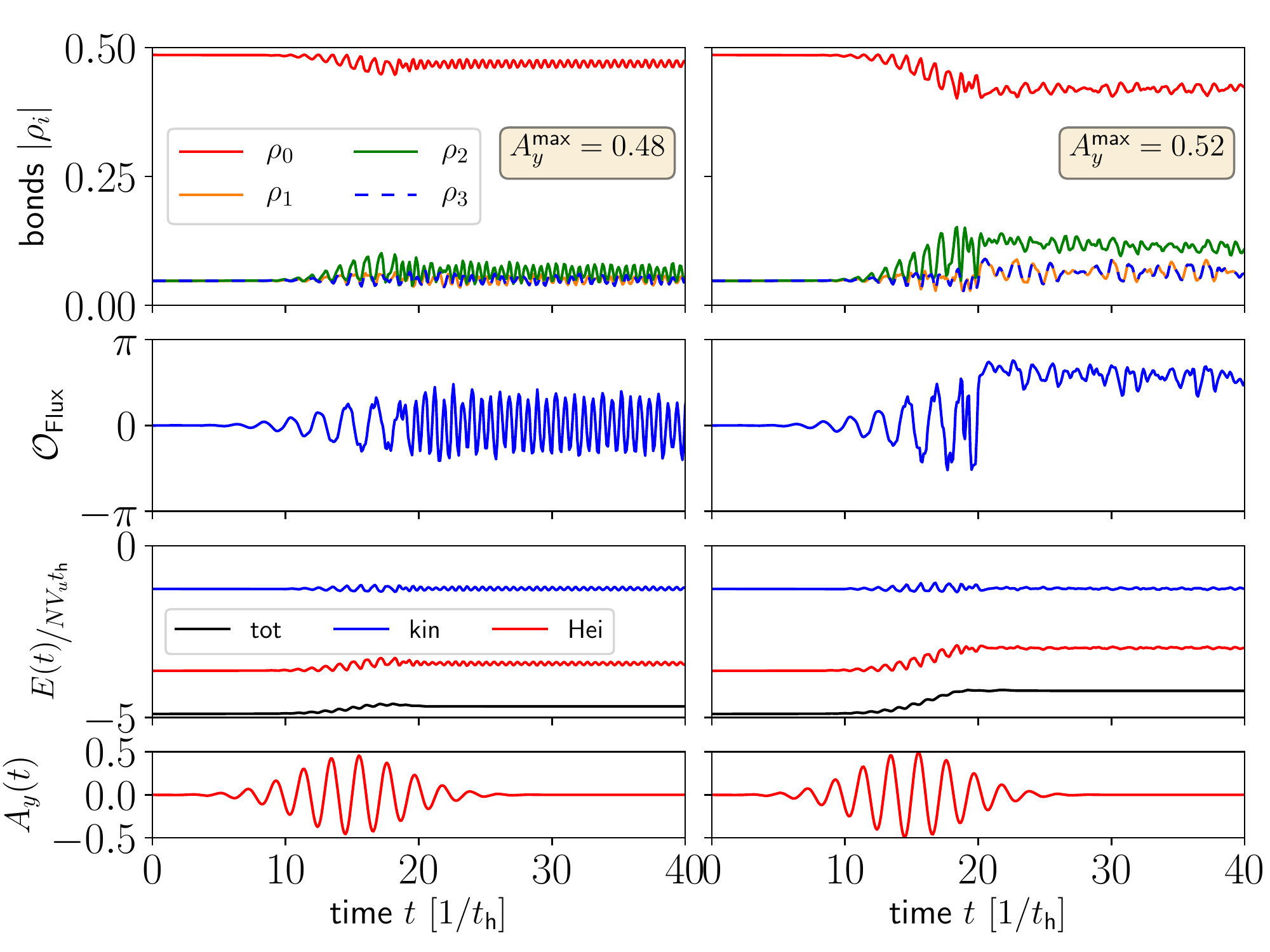}
 \caption{Mean-field dynamics of order parameters in the photoexcited Hubbard-Heisenberg model with $J = 15 t_\text{h}$, $U = 0$ and $V_u = 41 \times 41$ unit cells.
 Left: Pulse amplitude of $A_y = 0.48$ yields oscillations of $\mathcal{O}_\text{Flux}$ around average value zero. 
 Right: Pulse amplitude of $A_y = 0.52$ excites oscillations around non-vanishing average.
 \label{fig:mf_tevol_nnbonds_flux}}
\end{figure}

\begin{figure}
    \centering
    \includegraphics[width=0.5\textwidth]{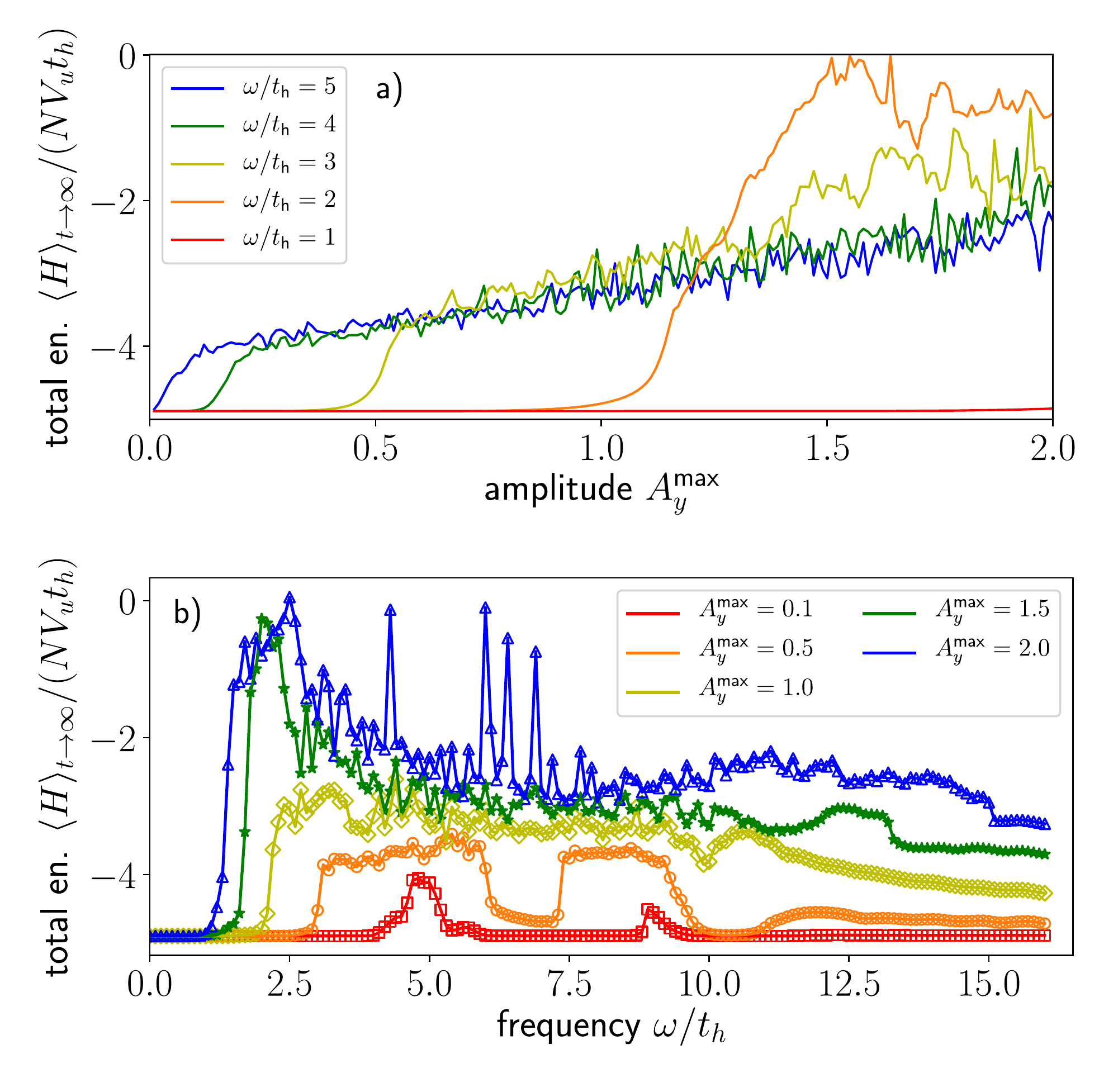}
    \caption{Energy absorption (mean-field dynamics) after a Gaussian pulse with $A_x = 0$, $A_y \neq 0$ in a Hubbard-Heisenberg square lattice with $V_u = 41 \times 41$ unit cells, $J/t_\text{h} = 15$ and $U = 0$.
    (a) post-pulse total energy (per unit cell) plotted against the pulse amplitude. Energy absorption is efficient only beyond a frequency-dependent threshold amplitude.
    (b) total energy plotted against the driving frequency for fixed values of the amplitude. At small amplitudes two absorption peaks are visible. }
    \label{fig:mf_ampl_freq_panel}
\end{figure}

Now we turn to the dynamics induced by a time-dependent electromagnetic field $E(t)$ using the Peierls substitution technique \cite{Peierls1933},
which describes a classical electromagnetic field (i.e. no quantized photons).
Within this approach the hopping matrix element acquires a time-dependent complex phase $t_\text{h} \text{e}^{i A_i(t)}, i \in \{ x, y \}$,
where $A_i(t)$ is the vector potential.
It is related to the electric field $E_i(t) = - \partial_t A_i(t)$ but for simplicity we concentrate the discussion on the vector potential itself with linearly polarized pulsed driving amplitudes of the form
\begin{equation}
 A_i(t) = A_i^\text{max} \text{e}^{-\frac{1}{2 \sigma^2} ( t - t_0 )^2} \sin\big( \omega(t - t_0) \big) .
 \label{eq:A_shape}
\end{equation}
We will always work at a fixed pulse width $\sigma = 4/t_\text{h}$.
In the two-dimensional model there is an additional freedom of choice for the polarization direction of the pulse.
For initial Peierls states there are two special directions: along or orthogonal to the strong bond.
We will mostly orient the vector potential along the direction of the strong bond ($A_x = 0, A_y \neq 0$) and for comparison orthogonal to it ($A_x \neq 0, A_y = 0$).

\subsection{Mean-field dynamics $N \rightarrow \infty$}

The simplest dynamical approach is mean-field theory.
It corresponds to setting the variance of the Gaussian Wigner function to zero and to consider only a single trajectory.
For now we set $U = 0$.
Fig.~\ref{fig:mf_tevol_nnbonds_flux} shows exemplary results for the order parameter dynamics with two different values of the driving amplitude $A_y$.
In both cases there are undamped coherent oscillations of the bonds and of $\mathcal{O}_\text{Flux}$ subsequent to the pulse.
It illustrates that within mean-field theory one cannot reach a stationary state due to the absence of dephasing and collisions.
Nevertheless, the system can absorb energy and it is possible to induce Flux order parameter oscillations around a non-vanishing average value.
However, the averaged absolute values of the bonds need not match with the saddle point symmetries of the equilibrium phases (e.g. all equal in the Flux phase) as can be seen in the right column.

Fig.~\ref{fig:mf_ampl_freq_panel} allows for a more systematic look at the energy absorption.
In (a) we vary the driving amplitude for some fixed values of the frequency $\omega$.
All curves in the figure display similar behavior:
There is almost no energy absorption up to a frequency-dependent threshold amplitude.
Above it, the total energy develops oscillatory patterns as a function of the driving amplitude.
We found that -- in contrast to the threshold regime -- the precise shape of these patterns can depend on the system size.
Nevertheless, the averaged trends in the data are consistent for different system sizes.
A way to make sense of this is to think about the Peierls phase as a minimum in a free energy landscape \cite{delaTorre2021} that is separated from other regions, e.g. the Flux phase minimum, by barriers.
If the system is only weakly excited, order parameter oscillations around the immediate vicinity of the Peierls phase minimum are induced.
At and beyond the threshold amplitude  other regions of the landscape become accessible.
The free energy \eqref{eq:free_en_temp} contains a discrete crystal momentum sum over the $\epsilon_k$,
which leads to an oscillatory fine structure of the mean-field free energy as a function of the bond operators and thereby likely affects the energy absorption if the system leaves the initial minimum.
In the finite-$N$ case, however, this fine structure will average out over many trajectories and so we do not expect the system size to play a significant role.
In Fig.~\ref{fig:mf_ampl_freq_panel}(b) the roles of frequency and amplitude are exchanged.
We observe two main peaks of the absorption at small amplitudes.
This is reminiscent of results reported for a driven non-interacting two-band model \cite{Shen2014b}.
The authors plot the occupation of the upper band against $\omega$ and observe a multi-peak structure.
In their paper the main peak position corresponded to the band gap and the existence of amplitude-dominated and frequency-dominated driving regimes are proposed.
It is not clear if similar arguments apply here since our band structure is not static and multiple phases with different single-particle spectra exist.
The gap in the single-particle energy spectrum of the Peierls phase shown in Fig.~\ref{fig:brillouin_disp} is larger than the peak positions observed in Fig.~\ref{fig:mf_ampl_freq_panel}(b).
However, the mean-field band structure does not describe single-particle excitations, which can well occur at a lower energy.
For driving amplitudes $A_y^\text{max} \gtrapprox 1$ the double peak structure disappears and some, sporadically large, oscillations occur.
These are again most likely due to the discrete $k$-space structure and will average out if multiple trajectories are used.



\begin{figure*}
	\centering
	\includegraphics[width=\textwidth]{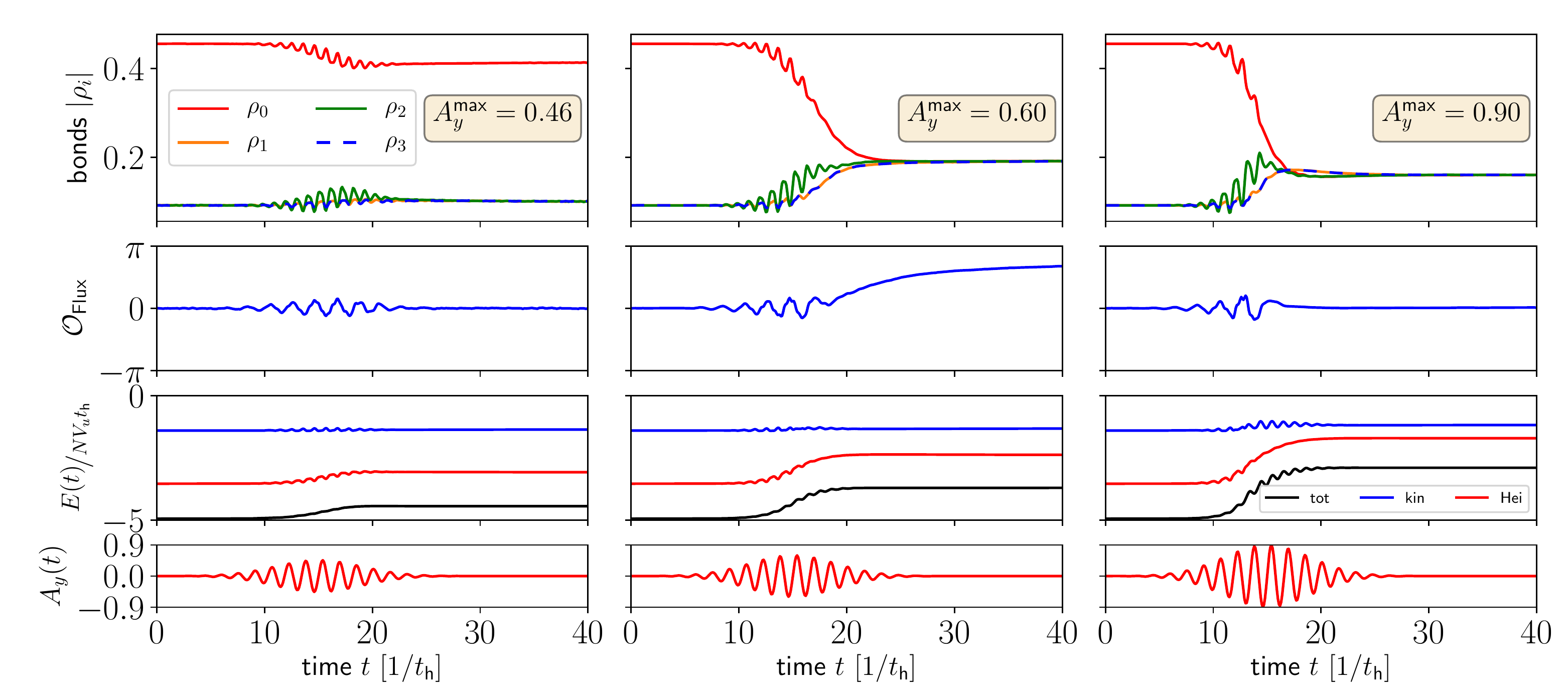}
	\caption{Dynamics of the order parameters during and subsequent to a photoexcitation in the Hubbard-Heisenberg model with $N = 300$ and for $V_u = 11 \times 11$ unit cells.
	The frequency of the sinusoidal drive is $\omega / t_\text{h} = 3$.
	The three columns represent three different maximum amplitudes of the vector potential $A_y(t)$ (i.e. fluences).
	The dynamics leads to transitions within the Peierls phase (left), from the Peierls phase to the Flux phase (middle) and to the uniform phase (right).}
	\label{fig:tevol_nnbonds_flux_freq3}
\end{figure*}

\subsection{Finite-$N$ model: mean-field + dephasing}

In this section we refine the previous discussion by including dephasing dynamics with fTWA.
This allows us to describe the formation of prethermal order during and subsequent to the pulse.
In the following we will always work on a lattice with periodic boundary conditions and $V_u = 11 \times 11$ unit cells, i.e. $V = 2 V_u = 242$ lattice sites.
We keep $J/t_\text{h} = 15$, $U = 0$
and prepare the system in the initial state \eqref{eq:ini_state_fin_N} as outlined in the Method section assuming that a stationary Peierls state is reached after a time of about $200 t_\text{h}^{-1}$ (cf. Fig.~\ref{fig:stat_order_params_vs_time_mult_N}).
We center the pulse at $t_0 = 250 t_\text{h}^{-1}$.
When we calculate the Flux order parameter $\mathcal{O}_\text{Flux}$ directly, we always turn on a weak symmetry breaking $\epsilon_{mn} = 10^{-3}$ (cf. Methods section). Peierls order parameters and current-current correlators are calculated without symmetry breaking field.
In Fig.~\ref{fig:tevol_nnbonds_flux_freq3} we show the time evolution of order parameters in an exemplary way for $\omega = 3 t_\text{h}$ and three values of the driving amplitude $A_y$ along the direction of the strong bond.
Time zero in the panels is set to $t_0 - 15 t_\text{h}^{-1}$.
In contrast to the $N \rightarrow \infty$ case, coherent order parameter oscillations subsequent to the pulse are damped out and a stationary state is reached.
Note that $| \rho_1 | = | \rho_3 |$, i.e. the spatial symmetry of the applied vector potential along the $y$-direction is preserved throughout the dynamics.
The three driving amplitudes lead to final states corresponding to the three equilibrium phases at half filling:
After weak driving the system remains in the Peierls phase with a smaller order parameter than in the initial state.
For intermediate amplitudes the Flux order parameter becomes non-vanishing and the system is Flux-ordered, while strong driving pushes the system into the Uniform phase.
The symmetries of the stationary state observables agree with the saddle point expectations.
We further observe that the post-pulse order parameter dynamics can continue even if the total energy is already at its stationary value.

\begin{figure*}
	\includegraphics[width=\textwidth]{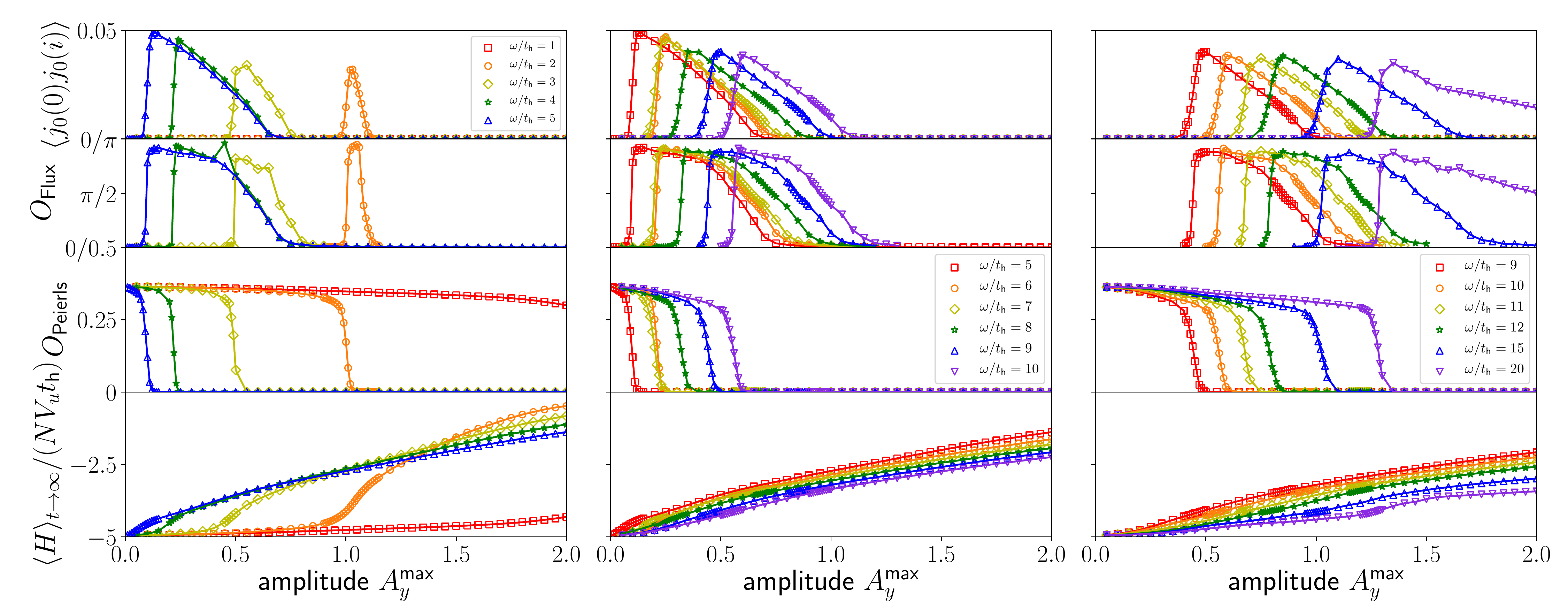}
	\caption{Late-time stationary values of the order parameters after photoexcitation with varying frequencies $\omega$ plotted against the amplitude of the vector potential $A_y^\text{max}$.
	The top row shows current-current correlators of $\rho_0$ for two unit cells with the largest spatial separation in the system (averaged over all unit cell pairs related by translational symmetry).
	The three columns correspond roughly to three different frequency regimes (see text).
	The data in this figure is obtained for a $11 \times 11$ unit cells Hubbard-Heisenberg model with $J = 15 t_\text{h}$ and $U = 0$.
	\label{fig:order_params_vs_fluence_freq}}
\end{figure*}

\begin{figure*}
	\includegraphics[width=\textwidth]{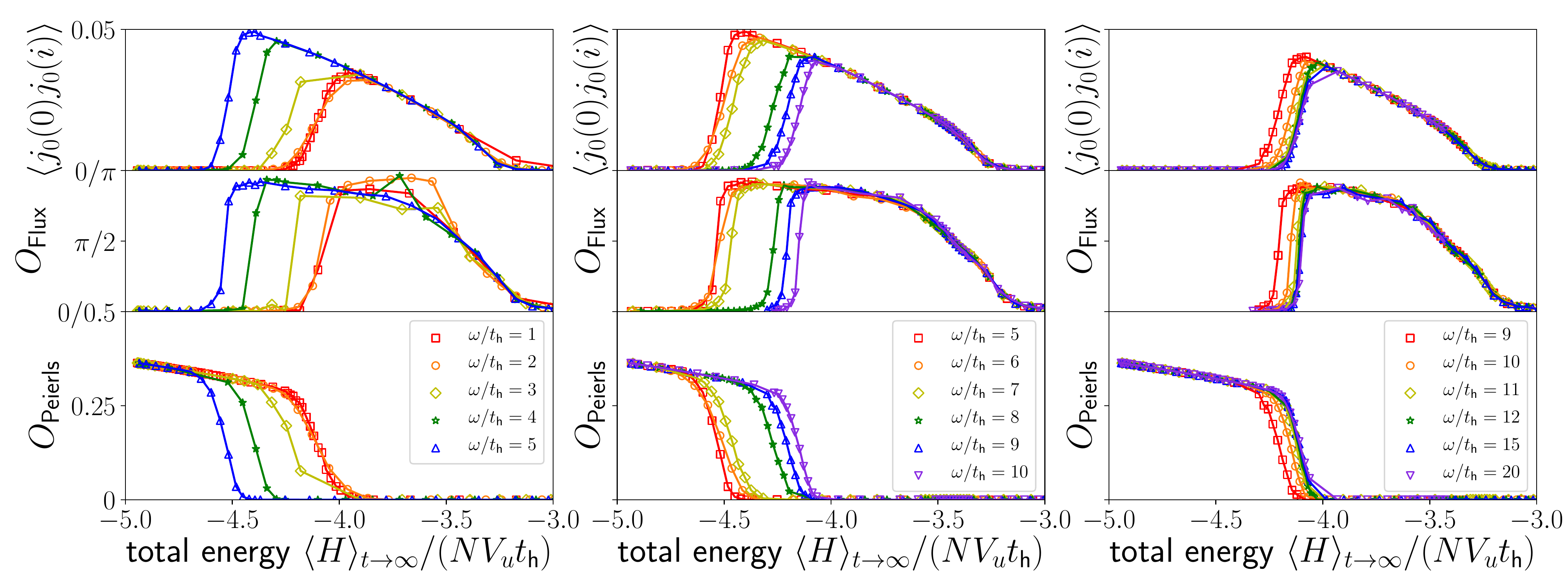}
	\caption{Same data as in Fig.~\ref{fig:order_params_vs_fluence_freq} but all final state order parameters are plotted against the total energy.
	In the first and third column the order parameters display a similar behavior for very low and very high frequencies.
	In the middle column there is a clear influence of the driving frequency on the order parameter characteristics.
	\label{fig:order_params_vs_en_freq}}
\end{figure*}

\subsection{Prethermal dependence on drive parameters}
In this section we study the influence of drive parameters like frequency, amplitude and polarization direction on the final state order parameters.
Fig.~\ref{fig:order_params_vs_fluence_freq} presents the numerical results as a function of $A_y$ for varying drive frequencies $\omega$.
In the uppermost row we plot the current-current correlation function for one of the bonds in a unit cell (the strong bond of the Peierls phase), which displays a sharp transition from Peierls to Flux and a broad transition range from Flux to Uniform order.
The Flux order parameter $\mathcal{O}_\text{Flux}$ agrees well with the correlator data except for deviations at low frequencies $\omega/t_\text{h} \leq 4$ and energies $-4 \leq E/t_\text{h} \leq -3.5$.
These deviations depend on the choice of the flux symmetry breaking strength and indicate that a slightly larger $\epsilon_{mn}$ might be needed in this regime.
Current-current correlators are thus the more robust quantifiers of Flux order.

One can roughly identify three frequency regimes, which correspond to the three columns.
At low values of the frequency $\omega \lessapprox 5 t_\text{h}$, there is a threshold amplitude -- analogous to the mean-field one but shifted to lower amplitudes -- at the transition from Peierls to Flux order.
The transition moves to lower amplitudes if the frequency increases.
For $\omega = 5 t_\text{h}$ the Peierls order parameter starts to decrease already at very low field amplitudes, which indicates that the drive is likely resonant with an elementary excitation of the system.
At higher frequencies, in the middle and right columns of the Figure, there is no threshold amplitude any more and the electrons absorb energy also for small values of $A_y$.
In these regimes, we find that the energy absorption decreases with increasing driving frequency.
This agrees with the physical expectation that energy absorption should be suppressed in the high-frequency regime due to the absence of available states for drive-induced transitions.
However, one cannot read off the elementary excitations of the system directly from the mean-field band structure in Fig.~\ref{fig:brillouin_disp}:
The Peierls phase has a large gap, while the Flux phase is gapless --  although thermodynamically they are almost degenerate.
One way to obtain a more detailed understanding of elementary excitations of the system would be to consider quantities like non-equilibrium spectral functions~\cite{Paeckel2020}, which is beyond the scope of this work.

\begin{figure}[ht]
	\includegraphics[width=0.5\textwidth]{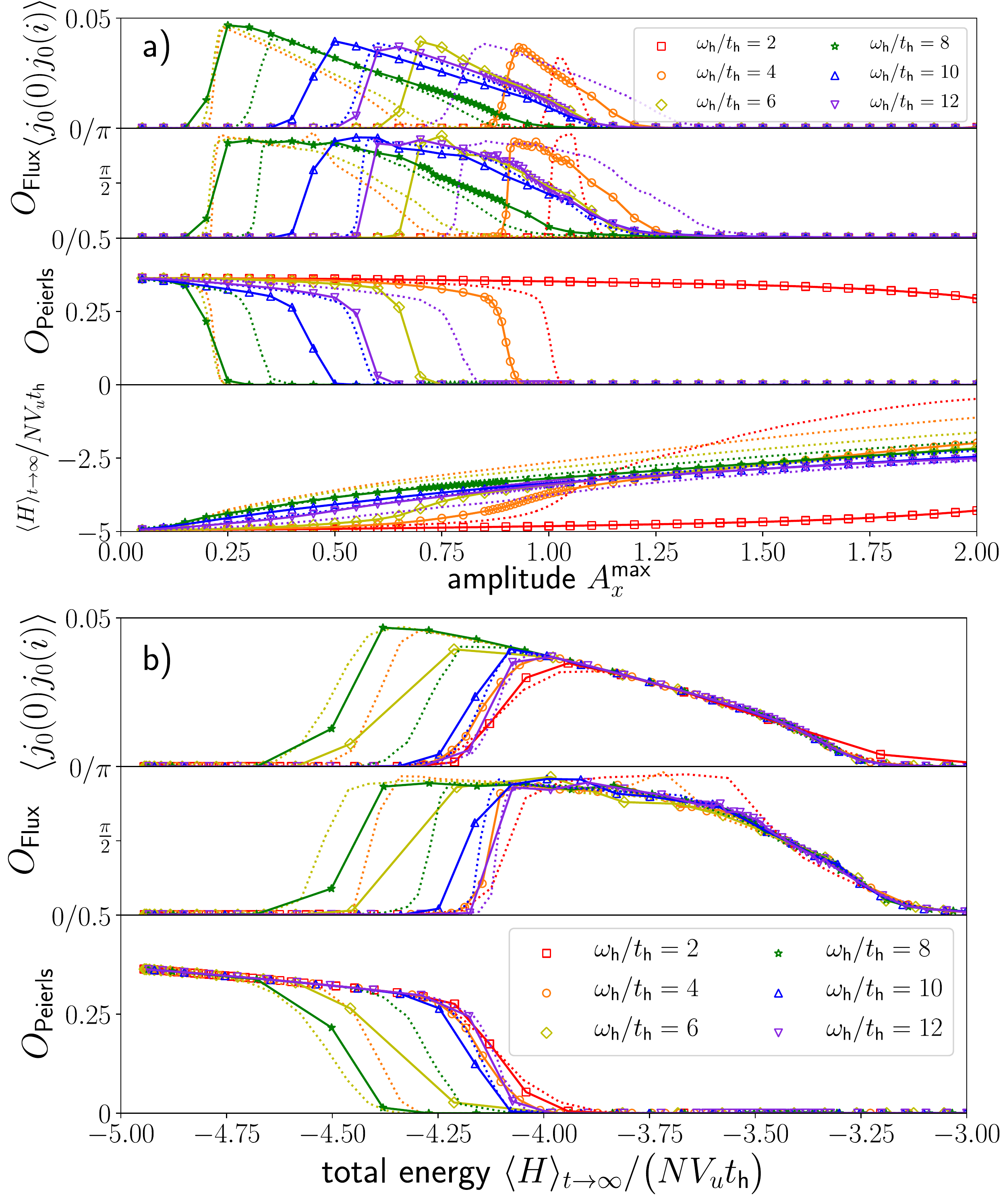}
	\caption{Late-time stationary values of the order parameters after a pulse with a vector potential that is polarized in the direction orthogonal to the strong bond of the Peierls phase ($A_x \neq 0$, $A_y = 0$) and for three different driving frequencies.
	The lattice has $V_u = 11 \times 11$ unit cells and $J/t_\text{h} = 15$, $U = 0$.
	\label{fig:panel_hom_Ax}}
\end{figure}

Naturally, a question raised by the stationary states of Fig.~\ref{fig:tevol_nnbonds_flux_freq3} is whether the post-pulse order could as well be a result of heating.
In particular, the mean-field finite-temperature equilibrium phases in Fig.~\ref{fig:eq_phases_temp_J15} follow the same sequence Peierls to Flux to Uniform as a function of temperature.
To shed more light on this question we plot the order parameters of Fig.~\ref{fig:order_params_vs_fluence_freq} directly against the total energy of the system after the pulse, shown in Fig.~\ref{fig:order_params_vs_en_freq} and again grouped by frequency.
At very low and high energies, the order parameters follow universal lines, independent of $\omega$.
The transition from Peierls to Flux order, in contrast, depends explicitly on the driving frequency.
The fact that not all curves lie on top of each other indicates the non-thermal nature of the stationary states.
Let us look at the regime of low driving frequencies.
As the total energy increases, the Peierls order parameter shrinks linearly down to a point where it decays and a non-zero Flux order parameter is found.
With increasing driving frequency this transition point shifts to lower energies.
In the intermediate frequency regime the transition moves back to higher energies.
In the high-frequency regime, finally, we find that for $12 t_\text{h} \lessapprox \omega$ the order parameter curves for all frequencies collapse.
The initial Peierls order is stable over a maximal energy range.


\begin{figure}[ht]
	\includegraphics[width=0.5\textwidth]{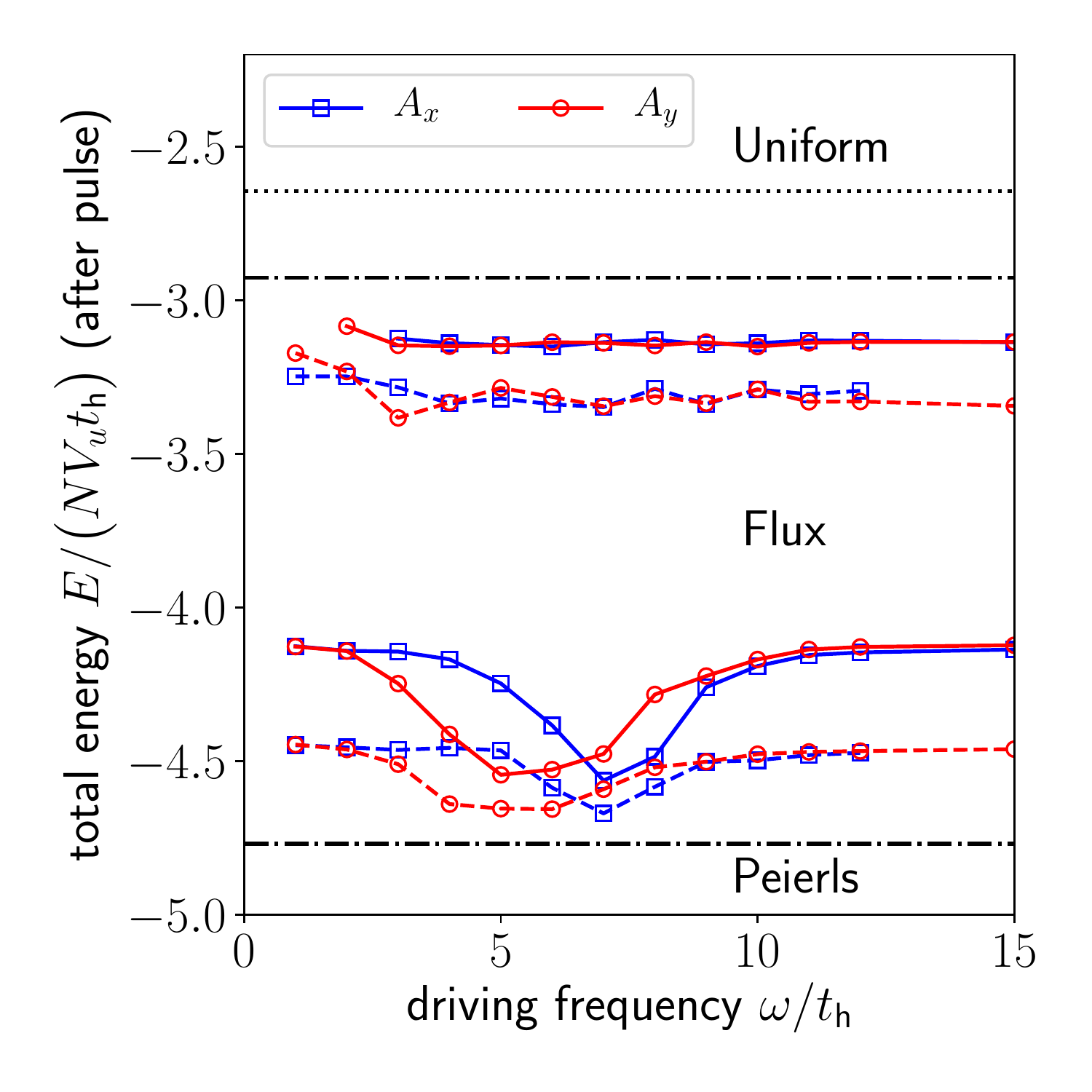}
	\caption{Transition energies between the three equilibrium phases in the Hubbard-Heisenberg model for $J/t_\text{h} = 15$ and $V_u = 11 \times 11$ unit cells after a pulse with driving frequency $\omega$.
	Red symbols correspond to pulses with $A_y = 0$, $A_x \neq 0$ (along the direction of the strong Peierls bond), blue symbols to the orthogonal direction.
	The dots connected with solid lines use $N = 300$, dashed lines $N = 250$.
	Black lines: mean-field model $N \rightarrow \infty$.
	Black dotted line: thermal transition from Flux to Uniform phase, 
	upper black dashdotted line: energy $E$ above which the Peierls saddle point disappears, lower black dashdotted line: energy of the Peierls state at which the first order transition to the Flux phase occurs.\label{fig:transition_en_vs_freq}}
\end{figure}

So far we discussed pulses with $A_x = 0$, $A_y \neq 0$, i.e. polarized linearly along the direction of the strong bond in a unit cell of the Peierls phase.
Let us compare the results to simulations with a vector potential along the $x$-direction.
Fig.~\ref{fig:panel_hom_Ax} shows results for the late-time stationary state value of the order parameters after a pulse with $A_x \neq 0$, $A_y = 0$ for three different values of the driving frequency $\omega$.
For frequencies $\omega \lessapprox 6 t_\text{h}$, the amount of absorbed energy as a function of $A_x^\text{max}$ is reduced compared to $A_y$-driving, in particular at high amplitudes.
The transition from Peierls to Flux order in Fig.~\ref{fig:panel_hom_Ax}(b) happens in a similar way to Fig.~\ref{fig:order_params_vs_en_freq}, although the regime with the earliest departure from Peierls order is shifted to higher frequencies.

Finally, we extract the transition energies from the order parameter curves in order to create the ``prethermal phase diagram'' in Fig.~\ref{fig:transition_en_vs_freq}.
At the Peierls-Flux transition we fit a sigmoid function to $\mathcal{O}_\text{Peierls}$ around the transition and determine the energy at the half-height sigmoid.
For the Flux-Uniform transition we extract the energy where the current-current correlator vanishes.
We show two sets of data: the one with $N = 300$ discussed so far and for comparison another set obtained with $N = 250$ (more detailed presentation in Appendix~\ref{app:data_N250}).
The transition Flux to Uniform is mostly independent of the driving frequency and field orientation except for a little uptrend at low frequencies,
which needs to be reexamined with more data points in the transition range.
In contrast, the Peierls to Flux transition is clearly dependent on the parameters of the drive.
We find a window of frequencies for which the transition occurs around a lowest energy.
It is shifted to higher frequencies in the case of $A_x$-polarization, which demonstrates the relevance of the spatial structure of order for optical excitations in an extended 2D system.
In this regime the transition occurs with the least amount of absorbed energy, i.e. avoiding additional heating.
At low and high frequencies the transition energies approach similar values.
It would be interesting to see if these limits coincided with the thermal transition of the system at finite $N$, which requires a more detailed knowledge of the thermal finite-$N$ phase diagram.
The $N = 250$ data contains the same qualitative signatures for the transition values but shifted to lower energies.
A comparison with the mean-field lines suggests a linear downshift of the transition energy and a more complex $N$-dependence of the Peierls-Flux transition.

\subsection{The role of $U$}
\begin{figure}
	\includegraphics[width=0.5\textwidth]{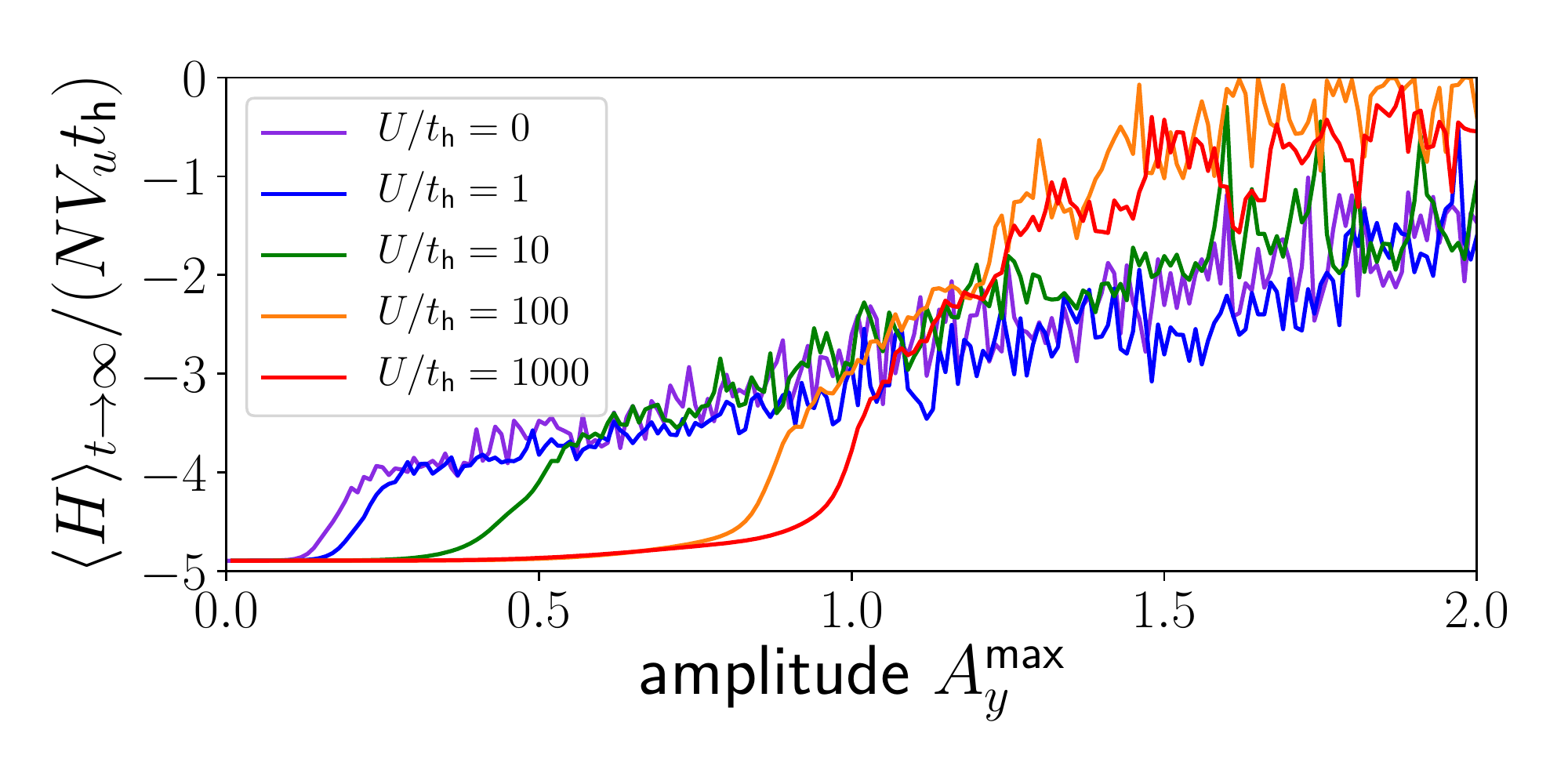}
	\caption{Total energy after a Gaussian pulse with frequency $\omega = 4 t_\text{h}$ as a function of the driving amplitude $A_y$ for some values of the Hubbard interaction strength $U$ ($J = 15 t_\text{h}$ is fixed). The data is obtained for the mean-field ($N \rightarrow \infty$) model on a lattice of $21 \times 21$ unit cells.\label{fig:en_vs_Ay_U}}
\end{figure}

Lastly, we would like to comment on the dynamical role of the Hubbard interaction $U$, which we omitted so far.
A non-vanishing value of $U$ will lead to a suppression of the on-site charge fluctuations during and subsequent to the pulse.
At half filling, $U/t_\text{h} \rightarrow \infty$ is the Heisenberg limit of the model.
To illustrate the effect of the $U$-term we show the post-pulse total energy in the $N \rightarrow \infty$ model for a driving frequency $\omega/t_\text{h} = 4$ in Fig.~\ref{fig:en_vs_Ay_U}.
The threshold amplitude for energy absorption is shifted to higher amplitudes upon increase of $U$.
However, at high driving amplitudes the system seems to heat up more than in the $U = 0$ case.
Without going into a detailed discussion of these effects, we can at least make the fundamental observation that parameters which are irrelevant in equilibrium (like $U$ in this case) may be relevant out-of-equilibrium.
In particular, one could design dynamical protocols to determine such parameters.
In this work we will not discuss the role of $U$ at finite values of $N$ because then the value of $U$ will also be relevant for the equilibrium phase diagram,
which goes beyond the scope of this paper.

\section{Conclusion and Discussion\label{sec:discussion}}

In this paper, we studied the photoinduced dynamics of order parameters in a two-dimensional interacting electron model with competing orders, driven by a pulsed sinusoidal vector potential.
A microscopic modeling of the dynamics on the mean-field level leads to undamped coherent oscillations that do not allow for a direct extraction of order parameter information.
We demonstrated that finite-$N$ corrections to the dynamics within the fTWA scheme include dephasing effects and lead to stationary states whose symmetries agree with those of the equilibrium phases.
Furthermore, we argued that the observed post-pulse order in the system cannot be thermal because otherwise the system should, for a given total energy $E$, approach a (unique) state with the lowest free energy.
In the numerical data, in contrast, we observed that the final stationary state also explicitly depends on drive parameters like frequency or field direction.
This dependence is particularly pronounced at intermediate frequencies,
while in the low and high frequency regimes the order parameter curves collapse as functions of the total energy.
Dephasing-induced prethermal order describes properties of an electron system on transient timescales before, in the full quantum time evolution, thermalization sets in.
The intermediate frequency regime is interesting because it allows to switch the order in the system (on this transient timescale) with a minimal amount of heating.
Therefore, it might be particularly relevant target for experiments since more energy in the system will suppress values of order parameters etc. and renders transient effects more difficult to observe.
Our findings motivate a search for similar frequency regimes in other models of competing ordered phases, for instance of superconductivity and charge density wave order.

In the main text we considered values of $N = 300$ and $N = 250$,
which are clearly off the conventional condensed matter case of $N = 2$.
Our choice of $N$ is limited in practice by our state preparation procedure, which starts from a mean-field state and then switches to finite $N$.
However, since we are interested in the leading order correction to the mean-field dynamics, a large value of $N$ is justified and convenient.
Contributions of order $1/N^2$ within fTWA need not agree with the correct quantum dynamics, as we demonstrated elsewhere by the example of the SU($N$)~Hubbard model~\cite{Osterkorn2020}.
Dynamical time scales and quantitative values of order parameters will depend on the chosen value of $N$ but within the range of validity of our theory we do expect to observe the same qualitative behavior also at smaller values of $N$.
The formalism developed here can directly be applied to other SU($N$) models.
Interesting candidates are, for instance, models with charge-density waves in equilibrium, like a SU($N$)~$t$-$V$ model.
While large-$N$ models provide a very natural field of application for semiclassical methods,
the fTWA method is not restricted to it and can be used to improve mean-field studies of more generic order parameter constellations, e.g. in the context of light-induced superconductivity~\cite{Sentef2017}.
The range of validity of fTWA, however, needs to be assessed carefully if no semiclassical expansion parameter is present.

To overcome the limitations of our initial state preparation scheme one could explicitly calculate $1/N$ corrections to the saddle points within field theory,
or try a calculation of fluctuations around the mean-field state within a self-consistent RPA~\cite{Jemai2005} or flavor wave~\cite{Kim2017a} theory.
Besides, efficient quantum Monte-Carlo~(QMC) codes~\cite{Assaad2022} exist for SU($N$) models,
which allow one to take the initial equilibrium state correlations directly from a QMC calculation.
Further method development should try to extend fTWA beyond prethermalization.
One way to generate real thermalization dynamics could be to take guiding from the BBGKY hierarchy~\cite{Lacroix2014,Pucci2016,PineiroOrioli2017} and add more dynamical variables to the equations of motion.
Another possible procedure would be to manually switch on the dynamics generated by a quantum Boltzmann equation (QBE) after prethermalization.
The QBE manifestly evolves the system towards a thermal fixed point and it could also allow for an estimate of lifetimes of prethermal states~\cite{Biebl2017}.

Finally, since the semiclassical dynamics does not require translational invariance of the lattice system,
a straightforward extension of the work presented here is to consider spatially local photoexcitations and to study the role of inhomogenities for the formation of (prethermal) order~\cite{Picano2021}.
Research in this direction is currently in progress.
Our framework also allows to prepare systems with boundaries between different mean-field-like phases to study the ordering dynamics at an interface microscopically~\cite{Sun2020}.

\begin{acknowledgments}
 We thank S.R.~Manmana, N.~B\"{o}lter, M.~Heyl and A.~Polkovnikov for useful discussions. 
This work is funded by the Deutsche Forschungsgemeinschaft (DFG, German Research Foundation) - 217133147/SFB 1073, project B07.
\end{acknowledgments}

\appendix
\section{Details on the determination of the mean-field states\label{app:details_mf}}

In the limit $N \rightarrow$ the model is effectively described by a Hartree mean-field theory,
which we will rederive in this appendix.
Let us consider the operators $\frac{1}{N} \sum_{\alpha = 1}^N \big( c_{i\alpha}^\dagger c_{j\alpha} - \frac{1}{2} \delta_{i,j} \big)$ and their deviations
\begin{equation}
\Delta\hat\rho_{ij} = \hat\rho_{ij} - \rho_{ij}
\label{eq:mf_deviation}
\end{equation}
from the expectation value $\rho_{ij} = \langle\hat\rho_{ij}\rangle$.
Plugging \eqref{eq:mf_deviation} into the model Hamiltonian
yields the following representation of the interaction terms
\begin{align}\begin{split}
\hat H_\text{Hei} &= -J N \sum_{\langle i,j \rangle} \big(\rho_{ij} + \Delta\hat\rho_{ij} \big) \big( \rho_{ji} + \Delta\hat\rho_{ji} \big) \\
&= -J N \sum_{\langle i,j \rangle} \big( | \rho_{ij} |^2 + \rho_{ij} \Delta\hat\rho_{ji} + \rho_{ji} \Delta\hat\rho_{ij} \big) \\
&\qquad + \mathcal{O}(\Delta\rho)^2 \\
&\simeq -J N \sum_{\langle i,j \rangle} \Big( - | \rho_{ij} |^2 \\ 
&\qquad\qquad + \frac{1}{N} \sum_{\alpha = 1}^N (\rho_{ij} c_{j\alpha}^\dagger c_{i\alpha} + \rho_{ji} c_{i\alpha}^\dagger c_{j\alpha} \big) \Big)
\end{split}\end{align}
\begin{align}\begin{split}
\hat H_\text{Hub} &= U N \sum_i \big( \rho_{ii} + \Delta\hat\rho_{ii} \big)^2 \\
&= U N \sum_i \big( |\rho_{ii}|^2 + 2 \rho_{ii} \Delta\hat\rho_{ii} \big) + \mathcal{O}(\Delta\rho)^2 \\
&\simeq U N \sum_i \Big( -|\rho_{ii}|^2 \\
&\qquad\qquad + \frac{2}{N} \sum_{\alpha = 1}^N \rho_{ii} \big( c_{i\alpha}^\dagger c_{i\alpha} - \frac{1}{2} \big) \Big)
\end{split}\end{align}
In the following we neglect the order $( \Delta\rho )^2$ terms.
In order to construct the ground state of the model we need to specify the lattice geometry and choose a unit cell.
Let us concentrate, for simplicity, on a one-dimensional model and give the generalization to two spatial dimensions in the end.
Here, we choose a unit cell with two sites $A$ and $B$.
We call the on-site elements of the $\rho$-variables $\rho_A := \rho_{i \in A, i \in A}$ and $\rho_B := \rho_{i \in B, i \in B}$.
The nearest neighbor bonds are called $\rho_0 := \rho_{i \in A, (i+1) \in B}$ and $\rho_1 := \rho_{i \in B, (i+1) \in A}$.

Next, we transform all non-local operators to momentum space.
A way to do this is to work directly in the reduced Brillouin zone $\text{rBZ} \subseteq (-\pi/2,\pi/2]$.
This corresponds to the following transformation rules with $Q = \pi$ (also described in Ref.~\cite{Shen2014})
\begin{align}\begin{split}
 c_{i \in A}^\dagger &= \frac{1}{\sqrt{V}} \sum_{k \in \text{rBZ}} \text{e}^{-i k r_i} \big( c_{k}^\dagger + c_{k + Q}^\dagger \big) \\
  c_{i \in B}^\dagger &= \frac{1}{\sqrt{V}} \sum_{k \in \text{rBZ}} \text{e}^{-i k r_i} \big( c_{k}^\dagger - c_{k + Q}^\dagger \big)
\end{split}\end{align}

Introducing $\epsilon_k := 2 t_\text{h} \cos(k)$ and $\chi_k := J \big( \rho_0 \text{e}^{-ik} + \rho_1^\ast \text{e}^{ik} \big)$ we obtain the following representation of the Hamiltonian

\begin{widetext}
\begin{equation}
 ( H ) = \begin{pmatrix} c_k^\dagger & c_{k + Q}^\dagger \end{pmatrix} \begin{pmatrix} \epsilon_k - \operatorname{Re}(\chi_k) + U (\rho_A + \rho_B) - \mu & i \operatorname{Im}(\chi_k) + U (\rho_B - \rho_A) \\ -i \operatorname{Im}(\chi_k) + U (\rho_B - \rho_A) & -\epsilon_k + \operatorname{Re}(\chi_k) + U(\rho_A + \rho_B) - \mu \end{pmatrix} \begin{pmatrix} c_k \\ c_{k+Q} \end{pmatrix}
\end{equation}
\end{widetext}

In this publication we only consider half filling, $\mu = U (\rho_A + \rho_B)$.
Diagonalization of the Hamiltonian yields the following set of eigenenergies
\begin{align}\begin{split}
 E_k &= \pm \Big\{ - \Big[ -\big( \epsilon_k - \operatorname{Re}(\chi_k) \big)^2 - \operatorname{Im}(\chi_k)^2 \\
 &\qquad\qquad - U^2 \big( \rho_B - \rho_A \big)^2 \Big] \Big\}^{1/2} \\
 &= \pm \Big\{ \big| \epsilon_k - \chi_k \big|^2 + U^2 (\rho_B - \rho_A)^2 \Big\}^{1/2} .
\end{split}\end{align}

In two spatial dimensions the procedure is analogous.
However, since we use the rotated unit cell depicted in Fig. \ref{fig:tilted_unit_cell},
the $k$-values are defined with respect to the tilted lattice of unit cells.
\begin{align}\begin{split}
 \epsilon_k &= 2 t_\text{h} \Big( \cos\Big( \frac{\sqrt{2}}{2} (k_x - k_y) \Big) \\
 &\qquad\quad + \cos\Big( \frac{\sqrt{2}}{2} (k_x + k_y) \Big) \Big), \\
 \chi_k &= J \Big( \rho_0^\ast \text{e}^{-i \frac{\sqrt{2}}{2}(k_x + k_y)} + \rho_1^\ast \text{e}^{i \frac{\sqrt{2}}{2} (k_x - k_y)} \\
 &\qquad\quad + \rho_2 \text{e}^{i \frac{\sqrt{2}}{2} (k_x + k_y)} + \rho_3 \text{e}^{-i \frac{\sqrt{2}}{2} (k_x - k_y)} \Big) .
\end{split}\end{align}

Rewritten in terms of the axes of the original lattice (here denoted $k'$) the quantities read as follows
\begin{align}\begin{split}
 \epsilon_k &= 2 t_\text{h} \big( \cos(k_x') + \cos(k_y') \big), \\
 \chi_k &= J \big( \rho_0^\ast \text{e}^{-i k_y'} + \rho_1^\ast \text{e}^{i k_x'} + \rho_2 \text{e}^{i k_y'} + \rho_3 \text{e}^{-i k_x'} \big) .
\end{split}\end{align}


\begin{figure}
	\includegraphics[width=0.5\textwidth]{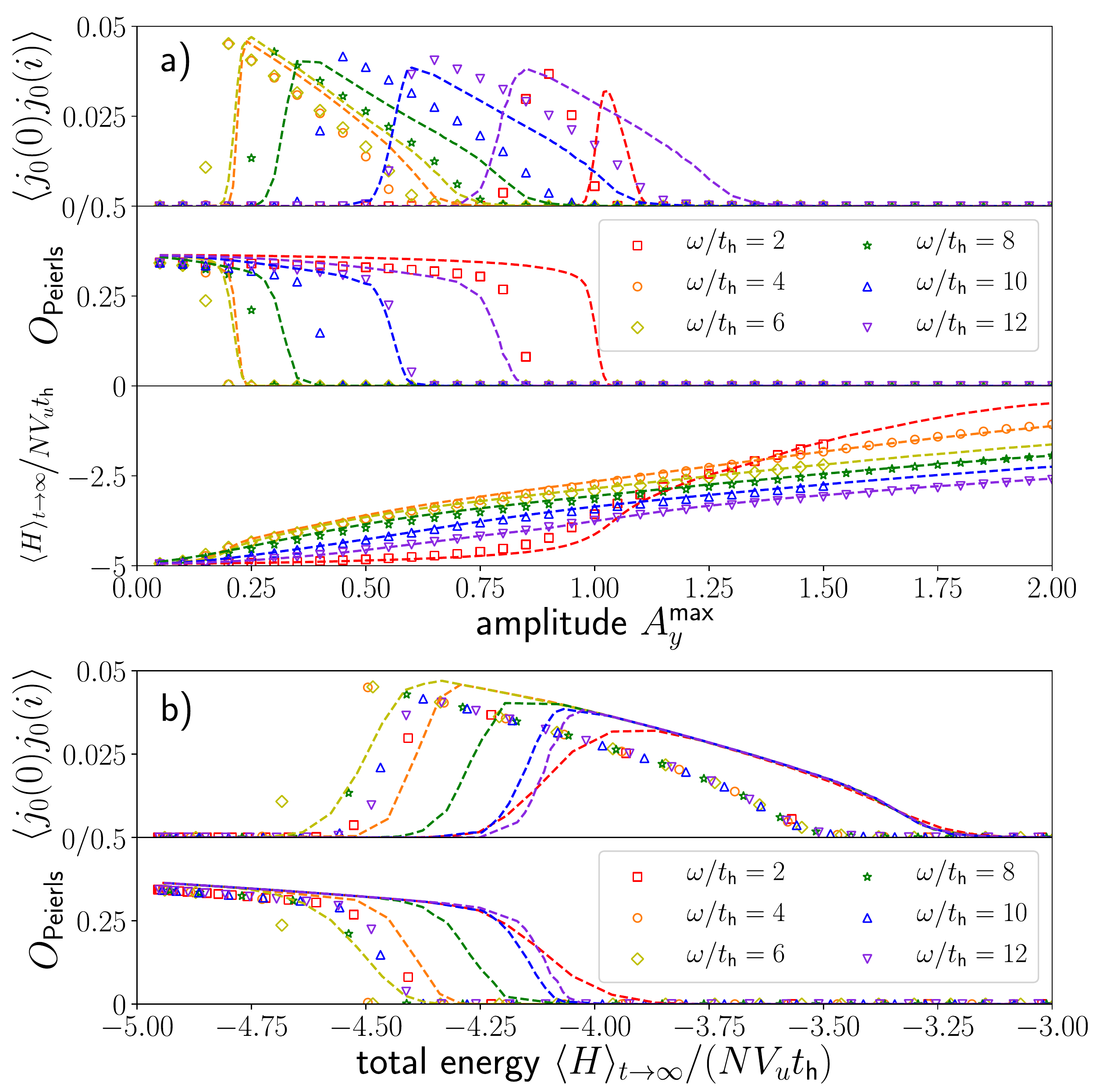}
	\caption{Order parameter dynamics for a Gaussian pulse ($A_x = 0$, $A_y \neq 0$) in the $N = 250$ Hubbard-Heisenberg model on a lattice with $V_u = 11 \times 11$ unit cells. The symbols represent $N = 250$ while the dashed line is the $N = 300$ case from the main text.
	\label{fig:panel_hom_Ay_n250}}
\end{figure}

\section{Order Parameter Data for $N = 250$\label{app:data_N250}}

In Fig.~\ref{fig:panel_hom_Ay_n250} we compare the order parameter data for $N = 250$ with $N = 300$.
The total amount of absorbed energy in Fig.~\ref{fig:panel_hom_Ay_n250}(a), divided by $N$, coincides with the $N =  300$ data for most of the considered driving amplitudes. 
This is due to the fact that in both cases the mean-field contribution to the absorbed energy is dominating.
In contrast,
the final state order parameter values differ more strongly.
Fig.~\ref{fig:panel_hom_Ay_n250}(b) shows that for $N = 250$ both transitions occur at lower energies than for $N = 300$.
This agrees with the intuitive expectation of an increased significance of quantum fluctuations at lower $N$, which destabilize mean-field order.
Still, the qualitative behavior of the order parameters is consistent.

\section{Details on the fTWA numerics and on the preparation of the Wigner function\label{app:wigner}}

In the numerics of the finite-$N$ model we average all observables over at least 200 trajectories.
In most cases this is already sufficient for a converged numerical result, i.e. the values do not change significantly upon increase of the number of averaged trajectories.
In some cases, in particular close to the Peierls-Flux transition, more samples are needed and we typically take into account at least 2000 trajectories.
For the solution of the ordinary differential equations we use the Runge-Kutta Cash-Karp error stepper from the \texttt{odeint} numerical library~\cite{Ahnert2011}.
Averaging over the trajectories is implemented using Welford's algorithm~\cite{Schubert2018}.
Tables \ref{tab:covariances} and \ref{tab:covariances_half_filling} collect the covariances that we used to set up the Gaussian Wigner function.
Table \ref{tab:covariances} shows the formulae for a general product state,
while Table \ref{tab:covariances_half_filling} is specific for the mean-field initial state used in the paper.

\begin{table*}
\caption{symmetrized covariance $\Gamma_{\alpha\beta,\mu\nu}^\text{symm} = \big\langle \frac{1}{2} \{ \hat \rho_{\alpha\beta}, \hat \rho_{\nu\mu} \} \big\rangle^\text{c.c.}$, symmetrized pseudo-covariance $C_{\alpha\beta,\mu\nu}^\text{symm} = \big\langle \frac{1}{2} \{ \hat \rho_{\alpha\beta}, \hat \rho_{\mu\nu} \} \big\rangle^\text{c.c.}$. Note that at half filling all ``$-$'' states are occupied ($n_{k-} = 1$) and all ``$+$'' states are unoccupied ($n_{k+} = 0$). Hence $n_{k\pm} + n_{l\pm} - 2 n_{k\pm} n_{l\pm} = 0$ and $n_{k\pm} + n_{l\mp} - 2 n_{k\pm} n_{l\mp} = 1$. \label{tab:covariances}}
\begin{ruledtabular}
\begin{tabular}{cccccc}
$\rho_{\alpha\beta}$ & $\mu_{\rho_{\alpha\beta}}$ & $\Gamma_{\alpha\beta,\alpha\beta}^\text{symm}$ & $C_{\alpha\beta,\alpha\beta}^\text{symm}$ & $\sigma^2_{\operatorname{Re}(\rho_{\alpha\beta})}$ & $\sigma^2_{\operatorname{Im}(\rho_{\alpha\beta})}$ \\
\colrule
$\rho_{k\pm, l\pm}$ & $\delta_{kl} \big( n_{k\pm} - \frac{1}{2} \big)$ & $\frac{1}{2N} \big( n_{k\pm} + n_{l\pm} - 2 n_{k\pm} n_{l\pm} \big)$ & $\frac{\delta_{kl}}{N} n_{k\pm} (1 - n_{k\pm})$ & $\frac{1}{4N} \big( 1 + \delta_{kl} \big) \big( n_{k\pm} + n_{l\pm} - 2 n_{k\pm} n_{l\pm} \big)$ & $\frac{1}{4N} \big( 1 - \delta_{kl} \big) \big( \dots \big)$ \\
$\rho_{k\pm, l\mp}$ & $0$ & $\frac{1}{2N} \big( n_{k\pm} + n_{l\mp} - 2 n_{k\pm} n_{l\mp} \big)$ & $0$ & $\frac{1}{4N} \big( n_{k\pm} + n_{l\mp} - 2 n_{k\pm} n_{l\mp} \big)$ & $\frac{1}{4N} \big( \dots \big)$ \\
\end{tabular}
\end{ruledtabular}
\end{table*}

\begin{table*}
\caption{Note that at half filling all ``$-$'' states are occupied ($n_{k-} = 1$) and all ``$+$'' states are unoccupied ($n_{k+} = 0$). Hence $n_{k\pm} + n_{l\pm} - 2 n_{k\pm} n_{l\pm} = 0$ and $n_{k\pm} + n_{l\mp} - 2 n_{k\pm} n_{l\mp} = 1$. \label{tab:covariances_half_filling}}
\begin{ruledtabular}
\begin{tabular}{cccccc}
$\rho_{\alpha\beta}$ & $\mu_{\rho_{\alpha\beta}}$ & $\Gamma_{\alpha\beta,\alpha\beta}^\text{symm}$ & $C_{\alpha\beta,\alpha\beta}^\text{symm}$ & $\sigma^2_{\operatorname{Re}(\rho_{\alpha\beta})}$ & $\sigma^2_{\operatorname{Im}(\rho_{\alpha\beta})}$ \\
\colrule
$\rho_{k\pm, l\pm}$ ($k = l$) & $\frac{1}{2}$ & $0$ & $0$ & $0$ & $0$ \\
$\rho_{k\pm, l\pm}$ ($k \neq l$) & $0$ & $0$ & $0$ & $0$ & $0$ \\
$\rho_{k\pm, l\mp}$ & $0$ & $\frac{1}{2N}$ & $0$ & $\frac{1}{4N}$ & $\frac{1}{4N}$ \\
\end{tabular}
\end{ruledtabular}
\end{table*}



\bibliography{main}

\end{document}